\documentclass[prb,amsmath,amssymb,12pt,aps,preprint,showkeys]{revtex4}

\newcommand{\doublespace}{\renewcommand{\baselinestretch}{2}\large\normalsize}

\newif\ifpdf\ifx\pdfoutput\undefined\pdffalse\else\pdfoutput=1\pdftrue\fi
\newcommand{\pdfgraphics}{\ifpdf\DeclareGraphicsExtensions{.pdf,.jpg}\else\fi}

\usepackage{graphicx}
\usepackage{dcolumn}
\usepackage{bm}

\input epsf

\def\be{\begin{equation}}
\def\ee{\end{equation}}
\def\ben{\begin{equation*}}
\def\een{\end{equation*}}

\begin{document}

\pdfgraphics


\title{Transient growth and instability in rotating boundary layers}

\author{Philip Yecko}
\email{yecko@columbia.edu}

\affiliation{Department of Physics and Astronomy, Arizona State University,
  Tempe, Arizona 85287-1504  USA}

\affiliation{Department of Astronomy, Columbia University, New York, New York 10027  USA}

\author{Maurice Rossi}
\affiliation{Laboratoire de Mod\'elisation en M\'ecanique, Universit\'e Pierre et Marie Curie (Paris VI),
4 Place Jussieu, 75252 Paris FRANCE}

\date{\today}

\doublespace

\begin{abstract}

\noindent
The three-dimensional temporal instability of rotating boundary layer
flows is investigated by computing classical normal modes as well as by evaluating
the transient growth of optimal disturbances.  The flows examined are
the rotating Blasius~(RB) and the rotating asymptotic suction layers~(RAS), with the
rotation axis normal to the basic flow plane.
In agreement with an inviscid criterion, streamwise unstable  modes are found in both flow cases
for anti-cyclonic rotation: at high Reynolds numbers, one obtains the Rossby number unstable range
$0<1/{\rm Ro}< 0.57$ for RB, or $0<1/{\rm Ro}<1$ for RAS.
Critical Reynolds  and Rossby numbers are also determined in both instances. Moreover
the dependence of transient growth with respect to  wavenumbers, Rossby and Reynolds numbers
is presented for both cyclonic and anti-cyclonic r\'egimes. In particular, the {\it peak}
transient growth  is computed for a wide  range of parameter values within the cyclonic regime and
is shown to be  reduced by rotation. A scaling analysis with respect to the Reynolds number
is  performed showing that the standard ${\rm Re}^2$ scaling is recovered
only at very weak rotation.
Optimal disturbances resemble oblique vortices. At weak rotation, they are almost
streamwise vortices  though their structure departs from the classical non-rotating case.
Strong rotation  imposes two-dimensionality  and the optimal disturbances
vary weakly in the spanwise direction and exhibit growth by the Orr mechanism.


\end{abstract}


\keywords{Instabilities: Boundary Layer, Boundary Layers: 3D,
Geophysical Flows: Rotating Flows
}


\maketitle

\clearpage
\newpage

\section{Introduction}


The standard development of boundary layers may be transformed by various
effects, the modification of boundary conditions
(riblets, compliant walls) or the presence of bulk forces such as
centrifugal \cite{DR} and/or Coriolis forces,~\cite{ZebBot} to name just a few.
In astrophysical, geophysical and technological flows, rotation through
the Coriolis effect plays a major role: technological examples
appear most notably for turbomachinery but also include rotary
atomization nozzles. \cite{Lefebvre}  Astrophysical examples include
the solar tachocline \cite{SpiegelZahn} and, on a larger scale,
accretion disk flow at the disk-star boundary layer. \cite{Frank}

For shear flows $(U(x,y),V(x,y),0)$, it is known that rotation along the
$z-$axis can produce an instability generating streamwise oriented vortices,
also known as longitudinal rolls. \cite{TrittonDavies}. This instability arises
due to the imbalance of pressure and Coriolis forces. \cite{DR}  Rotating
channel flows have been shown to exhibit this instability, in both theory and
experiment, \cite{Hart,LezJon,AlfPer} while rotating free shear
flows have been predicted to be likewise unstable. \cite{YFMR}

However, this modal instability is present only for a certain range of
rotations rates and depends on the direction of the frame rotation
vector (here the $z$-axis) with respect to the vorticity vector of the
base flow. When the latter quantities are parallel
(anti-parallel), the rotation is called cyclonic
(anti-cyclonic).  Cyclonic rotation and strong anti-cyclonic
rotation are found to stabilize normal modes. On the contrary, weak
anti-cyclonic rotation results in unstable streamwise vortices. All
these properties have been verified for rotating Poiseuille flow,
\cite{Hart,LezJon,AlfPer} rotating planar wakes and
rotating mixing layers. \cite{YFMR}  In addition, instability
thresholds have been determined in terms of Rossby and Reynolds
numbers (the Reynolds number ${\rm Re}$ and Rossby number ${\rm Ro}$
are defined explicitly in \S III(A)).  Such an instability mechanism
is fundamentally inviscid and is reduced by the presence of viscosity.
For instance, in rotating Poiseuille flow, \cite{LezJon} a
critical point was found at ${\rm Re_C}\approx 89$ and Rossby number
${\rm Ro}=1$.  Even though the boundary layer over a flat plate
continues to be an essential prototypical flow in the study of
transition to turbulence, rotating boundary layers have been less
extensively studied.  In one early study, \cite{PotCha} it was found
that rotating Blasius flow (denoted hereafter by RB) is destabilized
by weak anti-cyclonic rotation. However stabilization for stronger
anti-cyclonic rotation was not found since strong rotation rates were
not considered.  This study was limited to a few fixed values of
streamwise wavenumbers and no critical point was sought.  The first
purpose of the present work is to investigate the characteristics of
the unstable normal modes of rotational boundary layer flows by
emphasizing the ranges of Reynolds and Rossby numbers where the
rotational instability is active.

In non-rotating shear flows (plane Couette, Poiseuille, boundary layer
flows), nearly streamwise vortices have also been observed during the
transition to turbulence (note that no fundamental connection exists
between these two kind of streamwise vortices).  These   are
thought to be the result of a {\it transient growth} process that can
occur even when linear stability theory predicts that all modes are
stable.  The ability of disturbances in shear flows to grow
transiently was first proposed by Kelvin \cite{Kelvin} and explored by
Orr; \cite{Orr} later, Ellingson and Palm, \cite{EllPal} Landahl
\cite{Landahl} and Hultgren and Gustavsson \cite{HulGus} proposed
mechanisms for this growth. Transient growth theory has taken shape
more recently  see e.g. papers \onlinecite{BobBro,ButFar,TTRD,RedHen},
reviews, Refs. \onlinecite{Grossmann,Reshotko} and the monograph
\onlinecite{SchHen}.  For steady flows, transient growth can
be related to the non-orthogonality, with respect to a suitable energy
norm, of the eigenvectors of the linear evolution operator. This
feature permits certain initial conditions to experience relatively
large amplification $G(t)$ at intermediate times even for
asymptotically stable flows.  The most amplified transient
disturbances  for boundary layers \cite{ButFar} typically take the form of streamwise
vortices. The well known streaks seen in  such flows at high
levels of free stream turbulence, are now thought to be formed as these
optimal vortices lift up \cite{Landahl} low velocity fluid from the
wall region and push down high velocity fluid toward the wall.
\cite{MatAlf2001}  In rotating boundary layers, there exists also a wide range of
rotation rates for which the   instability is not present and little
attention has been paid to this rotationally stable regime.  In
particular, the possibility that disturbances experience significant
{\it transient growth} in rotating boundary layers, has not been
addressed yet (although some preliminary results for rotating channel
flow appears in Ref.~\onlinecite{SchHen}). The second and major purpose of this
paper is to study transient growth in rotating boundary layer flows,
primarily in the regime where no unstable normal modes are present.  Transient growth factors, maximized in time, are
calculated for a broad range of wavenumbers and Rossby numbers.  Peak
transient growth factors, maximized both in time and in wavenumber
space, are also computed as a function of rotation.  Calculations based
on RB flow are presented alongside equivalent calculations based on
the rotating asymptotic suction~(denoted hereafter by RAS) profile. The base flows are discussed in
\S II, then in \S III the disturbance equations and calculation
methods are presented. Results for modal instability and transient
growth for both base flows are contained in \S IV and the study is
concluded in \S V.

\section{Base Flows}

To examine the effect of rotation on boundary layers, we choose as base flows
two well-known boundary layer solutions which are rapidly recalled  in the following sub-section:
the Blasius boundary layer and the asymptotic suction profile.

\subsection{Base Flows without rotation}

The paradigmatic  Blasius  velocity profile $U(y)$ \cite{SchGer} is related to the
streamfunction $f$ by $U(y) = U^* f'(\eta)$ where $U^*$ denotes the free-stream velocity, primes the
differentiation with respect to the similarity variable $\eta$. More precisely
$f$ and $\eta $  satisfy
\be
f''' + {1\over 2}f f'' = 0 ~\hbox{with}~\eta=y/\Delta(x)~\hbox{and}~\Delta(x)=\sqrt{\nu x / U^*}\;.
\label{blasius}
\ee
The quantity $\Delta(x)$ denotes  a boundary layer thickness defined in terms of the free-stream velocity
$U^*$, the downstream distance $x$ and the kinematic viscosity $\nu$.
Equation (\ref{blasius}) is easily evaluated using
straightforward numerical schemes (Falkner-Skan
routines built into {\sc Matlab} \cite{Matlab} were used in this work).
Blasius flow includes as well a weak transverse velocity,
\be
V(y) = {1\over 2} \sqrt{\nu U^*\over x} \left(\eta f' - f\right) ,
\ee
which is zero at the wall and approaches the constant
displacement velocity, $V^* = 0.8604 U^* \sqrt{\nu/x U^*}$
as $y,\eta\rightarrow\infty$.
The displacement thickness
\be
\delta_B = \int_0^\infty \left({1-{U\over U^*}}\right)dy
= \Delta(x)\lim_{\eta\rightarrow\infty}[\eta-f(\eta)]
= 1.7208~\Delta(x) \; ,
\ee
defines the length scale used  to build a Reynolds number
${\rm Re} = U^*\delta_B/\nu$.

The non-rotating asymptotic suction profile~(denoted hereafter by AS) describes the velocity field obtained when a  fluid flows
parallel to a surface on which uniform suction $V^*$ is applied. This boundary layer profile
\be
U(y)=U^*\left(1-e^{-y/\delta_S}\right);~~V(y)=-V^*~~\hbox{with}~\delta_S = {\nu\over V^*}
\label{ASU}
\ee
is a true Navier-Stokes solution and, contrary to  Blasius flow,  does not vary in the downstream direction.
The linear \cite{SchGer,HugRei} and nonlinear \cite{Hocking}
stability properties of flow (\ref{ASU}) have been long known: AS flow remains stable for much higher
Reynolds numbers than Blasius flow for a Reynolds number based on the displacement thickness  (for AS equal to
$\delta_S$) and free stream velocity, i.e. ${\rm Re} = U^*\delta_S/ \nu$. Note that the non-dimensional
transverse velocity is such that $V^*/U^* = 1/{\rm Re}$.

\subsection{Base Flows with rotation}

Let us now consider  a  frame rotating along the $z$-direction with a constant
angular velocity  $\Omega$ (see~Fig.1). By applying the usual boundary layer scalings \cite{SchGer} to the rotating Navier-Stokes equations,
the  nondimensional equations
read  in the limit of large ${\rm Re}$,
\be
{\partial u \over \partial x}+{\partial v \over \partial y }=0
\ee
\be
 u {\partial u \over\partial x}+v{\partial u\over\partial y}=
-{\partial p\over\partial x} + {\partial^2 u\over\partial y^2}+ {1\over {\rm Re}{\rm Ro}}v
\label{swmom}
\ee
\be
0=-{\partial p\over\partial y}-{1\over {\rm Ro}}u
\ee
where the Rossby number is defined such as ${\rm Ro} = U^*/2 \Omega \delta$
where $\delta $ stands for the displacement thickness~($\delta =\delta_B$ or
$\delta_S$).  Unless rotation is very strong (i.e. the rotation number 
${\rm Rot}= (2{\rm Ro})^{-1} >> 1$), the term ${1\over {\rm Re}{\rm Ro}}v$ is weak
relative to the leading order terms (the strongest rotation numbers used in
this study are ${\cal O}(1)$) and can thus be neglected in the streamwise
momentum equation, (\ref{swmom}). A non-rotating boundary layer profile such as the Blasius
profile or AS profile, is then preserved in the presence of rotation
\cite{PotCha} although rotation calls for an added cross-stream base pressure
gradient, $dP/dy = -{1\over {\rm Ro}}U(y)$ to balance the Coriolis force
resulting from the streamwise velocity $U(y)$.  The rotating asymptotic
suction (RAS) profile continues to be an exact solution of the rotating
Navier-Stokes equations if, in addition to the normal base pressure gradient,
$dP/dy = -{1\over {\rm Ro}}U(y)$, a streamwise pressure gradient $dP/dx =
{1\over {\rm Re}{\rm Ro}}V$ is included in the base flow to balance the
Coriolis term in the streamwise momentum equation.

In the following stability and transient growth calculations, the downstream
variation of boundary layers is not accounted for. Two major hypothesis
follow: (a) the streamwise profile is considered at a given streamwise position
$x=x_0$, and (b) the transverse velocity component is discarded, i.e. we set
$V=0$.  For all RB computations and most RAS calculations (unless otherwise
stated), the approximated profile ${(U(y),0,0)}$ will be hence used.  However,
as mentioned above, RAS flow with $V \ne 0$ is an exact parallel solution of
the Navier-Stokes equations; it can thus be used as a test for the validity of the
second hypothesis.  In the Appendix, a comparison between the two
possibilities $V \ne 0$ and $V=0$ is presented for RAS, based on growth rates
and neutral curves of normal modes as well as transient growth.  This study
shows that the effect of the transverse velocity on the normal modes and
transient properties is negligible.

\section{Mathematical Formulation}

\subsection{Linearized disturbance equations}

In a  frame rotating along the $z$-direction with a constant
angular velocity  $\Omega$ (see~Fig.1), the  dimensionless   Navier-Stokes Equations read 
\begin{equation}
{\partial{\bf u}_{tot}\over \partial t}
+ {\bf u}_{tot}\cdot\nabla{\bf u}_{tot}
+ {1\over{\rm Ro}}{\hat z}\times{\bf u}_{tot}
= - \nabla p_{tot} + {1\over {\rm Re}} \nabla^2{\bf u}_{tot};~~~\nabla\cdot{\bf u}_{tot} =0\
\label{NDNS}
\end{equation}
where non-dimensionalization has been performed using
free stream velocity  $U^*$, displacement thickness  $\delta$~($=\delta_B$ or $\delta_S$),
and pressure scale $\rho(U^*)^2$. Two dimensionless number then arise:  the Reynolds number
${\rm Re} = U^*\delta/\nu$ , and the Rossby number
${\rm Ro} = U^*/2 \Omega \delta$.

Following standard methods, the base state velocity \cite{footnote}
$(U(y),V,0) $ and pressure $P(y)$ are  perturbed by adding an infinitesimal disturbance
$(u_p,v_p,w_p)$ and $p_p$. These linear perturbations
can be  assumed to satisfy a  modal behavior in $x$,$z$,   viz.
\begin{equation}
(u_p,v_p,w_p;p_p) = ({u}(y,t), { v}(y,t),{ w}(y,t);{ p}(y,t)) e^{i(\alpha x + \beta z)} \; ,
\end{equation}
where  $\alpha,\beta$ are assumed to be purely real. By using
normal velocity  ${ v}(y,t)$ and normal vorticity
\be
{ \zeta}(y,t) = i\beta { u}  - i\alpha { w} ,
\ee
the linear  dynamics is described by one equation governing normal velocity~${ v}$:
\begin{widetext}
\begin{equation}
( \frac{\partial }{\partial t} + i\alpha U + V D)(D^2-k^2){ v} - i\alpha (D^2 U) { v}
+ {1\over{\rm Ro}}i\beta{\zeta} - {1\over{\rm Re}}(D^2-k^2)^2{ v}=0 ,
\label{OStime}
\end{equation}
\end{widetext}
and a second equation governing  normal vorticity~${ \zeta}$:
\begin{equation}
(\frac{\partial }{\partial t} + i\alpha U + V D ){ \zeta}
-{1\over{\rm Ro}}i\beta{ v} + i\beta (DU){ v}
- {1\over{\rm Re}}(D^2-k^2){ \zeta}=0 ,
\label{SQtime}
\end{equation}
where $k^2 \equiv \alpha^2+\beta^2$, $D \equiv d/dy$. Equation (\ref{OStime})~(resp. (\ref{SQtime})) is nothing but
 the Orr-Sommerfeld (resp. Squire)
equation  modified to account for   rotation.   Boundary conditions
\begin{equation}
{ v} = D{ v} = { \zeta} = 0 \;\;\;\;
\label{BCs}
\end{equation}
should be imposed  at the wall $(y=0)$ and in the far field $(y\rightarrow\infty)$. A dynamical system is thus defined for
${\bf{q}(y,t)}=( { v}(y,t),{\zeta}(y,t))^T$.

\subsection{Normal mode approach}

In this work, we study the above problem using two different approaches: the normal mode  approach
(cf asymptotic dynamics) and
the non-normal  approach (cf transient  dynamics).
The first approach consists in recasting the problem in \S III(A)   into  an eigenvalue problem by
assuming ${\bf q}={\bf{\tilde q}(y)}e^{-i\omega t}$ where ${\bf{\tilde q}}=( {\tilde v},{\tilde\zeta})^T$ and
$\omega=\omega_r+i\omega_i$ are complex. Clearly a mode having $\omega_i>0$ is  unstable.
Equations (\ref{OStime})-(\ref{SQtime}) can then be written
\begin{equation}
i\omega{\bf M{\tilde q}} = {\bf L{\tilde q}}
\label{mateval}
\end{equation}
where
\begin{gather}\label{defML}
{\bf M} = \begin{pmatrix} D^2-k^2 & 0 \\ 0 & 1 \end{pmatrix} \;\;
,\;\;
{\bf L} = \begin{pmatrix} {\cal L}_{OS} & {\cal R}_{OS}
\\ {\cal R}_{SQ} & {\cal L}_{SQ} \end{pmatrix}
\end{gather}
with
\be {\cal L}_{OS} = (i\alpha U + VD)
(D^2-k^2) - i\alpha (D^2 U)  - {1\over{\rm Re}}(D^2-k^2)^2 ,
\ee
\be
{\cal L}_{SQ} = i\alpha U + VD -{1\over{\rm Re}}(D^2-k^2) ,
\ee
\be {\cal R}_{OS} =
{1\over{\rm Ro}}i\beta ,~~{\cal R}_{SQ} = i\beta (DU)-{1\over{\rm Ro}}i\beta
\ee
It should be emphasized that as a result of rotation, the Orr-Sommerfeld equation is explicitly coupled
to the Squire equation  since both ${\cal R}_{OS}$ and ${\cal R}_{SQ}$ are nonzero
while only the Squire equation is coupled through ${\cal R}_{SQ}$  in the absence of
rotation. These equations together with the boundary conditions
\begin{equation}
{\tilde v} = D{ \tilde v} = { \tilde \zeta} = 0 \;\;\;\;
\label{BCstilde}
\end{equation}
imposed at the wall $(y=0)$ and in the far field $(y\rightarrow\infty)$
define the eigenvalue problem to be solved.  A Chebyshev collocation algorithm developed in previous work
\cite{YZF} is implemented. First the semi-infinite domain is approximated by a
sufficiently large interval $[0,L]$, a method known to be effective \cite{ButFar}.
In practice, the domain extended to sixteen displacement thicknesses $\delta$,
$L = 16~\delta$. This spatial interval  is then mapped to the interval $[-1, 1]$. Finally
eigenfunctions are  expanded into Chebyshev polynomials.
The number $N$ of polynomials  was generally $N=75$, although as
many as $N=125$ were used to give improved accuracy in some calculations.
Collocation points are the standard Gauss-Lobatto points.
The eigenvalue problem (\ref{mateval}) and boundary conditions
(\ref{BCstilde}) are then transformed into a corresponding $2N\times 2N$
matrix problem. To facilitate calculation, the eigenvalue problem
(\ref{mateval}) is actually written as
\be
i\omega{\bf {\tilde q}} = {\bf {\cal L}{\tilde q}}~~\hbox{with  }~~{\bf {\cal L}} \equiv {\bf M}^{-1}{\bf L}\;.
\label{dfL1}
\ee

\subsection{Transient Growth Calculations}

Transient amplification calculations have also been implemented on the problem in \S III(A)  according to the
singular value decomposition (SVD) method given by Reddy and
Henningson \cite{RedHen} and the algorithms published in Ref.~\onlinecite{SchHen}. The main steps of
the method are briefly described below; more details can be found in the references.
First, the standard kinetic energy norm
\be
||{\bf q(t)}||^2_E = \int_0^L (|Dv|^2 + k^2|v|^2 + |\zeta|^2) dy
\label{Enorm}
\ee
is used to measure the magnitude of a disturbance ${\bf q(t)}$ at time $t$.
At time $t$, one can define
\begin{equation}
G(t) = \sup_{{\bf q}(0)\neq 0}
 ~{||{\bf q(t)}||^2_E \over ||{\bf q(0)}||^2_E}
 = ||e^{i{\cal L}t}||^2_E \; ,
\label{Gt}
\end{equation}
where ${\bf q(0)}$ is an initial
disturbance and ${\cal L}$ is the linear operator defined in (\ref{dfL1}).
This quantity $G(t)$ represents the maximum possible energy amplification at time
$t$, optimized over all possible initial disturbances.
Finally the maximum or optimal growth  is defined as $G_{MAX}=\sup_{t\ge 0} G(t)$.
The  peak value  $ G_P({\rm Ro},{\rm Re}) \equiv \sup_{\alpha,\beta}
G_{MAX}(\alpha,\beta,{\rm Ro},{\rm Re})$  can also be   computed. Note that
$G_{MAX}$ is associated with a particular initial disturbance which
reaches $G_{MAX}$ at a specific  time $t_{MAX}(\alpha,\beta,{\rm Ro},{\rm Re})$  such that $G_{MAX}=G(t_{MAX})$.
Similarly one may define a specific  time $t_p({\rm Ro},{\rm Re})$ related to  $G_P$.

The quantity $G(t)$ is obtained as follows. Assume that the eigenvalues of (\ref{dfL1})-(\ref{BCstilde})
 are  sorted in order of
decreasing imaginary part $\omega_i$. One can approximate $G(t)$ by computing
the maximum possible energy amplification at time $t$, optimized over all possible initial combinations of
the $K$ eigenfunctions associated with the first $K$ eigenvalues of (\ref{dfL1}). In this case, it is possible
to transform the energy norm of the matrix exponential  (\ref{Gt}) to an ordinary
2-norm via  the following equation
\be
||e^{i{\cal L}t}||^2_E \approx ||{\bf F} e^{-i\Lambda
t}{\bf F^{-1}}||^2 \; ,
\label{2norm}
\ee
where $\Lambda$ is a $K\times K$ diagonal matrix with the first
$K$ eigenvalues ${\omega_1, \omega_2, ..., \omega_K}$ along the
diagonal and ${\bf F}$ is obtained by
Cholesky factorization of $K\times K$ Hermitian
matrix ${\bf {\cal M}}={\bf F^HF}$ which is  calculated using the inner product
of the $K$ eigenfunctions of (\ref{dfL1}), the inner product used being  the same as the one
defining the norm (\ref{Enorm}).

The approximation of (\ref{2norm}) is a result of the finite number $K$
used in the expansion. Following Ref~\onlinecite{RSH}, $K$ is chosen such that convergence is reached;
in practice, $K\approx 60$ was found to be sufficient.
The 2-norm on the RHS of (\ref{2norm}) can now be  evaluated using
SVD. This procedure  gives both $G_{MAX}$ and the $K$ expansion coefficients of the
disturbance associated with $G_{MAX}$ \cite{RedHen,ButFar}.
Note that, since eigenfunctions have been expressed as an expansion in
Chebyshev polynomials, all calculations are ultimately performed in
terms of the Chebyshev expansion coefficients. In particular, the energy norm
(\ref{Enorm}) and the weight matrix ${\bf {\cal M}}$ are easily recast in terms
of the expansion coefficients using the properties of Chebyshev
polynomials and their derivatives (for details see Ref.~\onlinecite{RSH}).

\section{RESULTS}

\subsection{Modal Stability Properties}

It can be shown that within inviscid theory, \cite{Hart,LezJon} the
stability of a rotating shear flow is analogous to that
of Boussinesq convection in which the total base vorticity $(2\Omega-dU/dy)$
plays the role of temperature gradient. The  dimensionless
instability criterion then  states that  total vorticity must be negative
somewhere in the flow \cite{Hart} i.e.
\be
\label{vortcond}
{\rm Ro}~{dU\over dy}>1 \; .
\ee
The relative vorticity of the base flow $-dU/dy$ which is directed along the $z$ direction,
is always negative for both
RB and RAS flows. Positive Rossby ${\rm Ro}>0$, thus
corresponds to oppositely directed rotation and flow vorticity, that is why it
is referred to, in the geophysical literature, as anti-cyclonic. This is the region we will consider  here since
negative Rossby ${\rm Ro}<0$ correspond to cyclonic case where this instability disappears.

\smallskip

Neutral stability curves of RB flow at
${\rm Re}=600$ and several Rossby numbers are shown in Fig. 2a. The region identified as ``TS''
corresponds to the (rotationally modified) Tollmien-Schlichting modes
of the non-rotating Blasius flow. The TS region is only slightly
modified by rotation, becoming discernibly larger  only when ${\rm Ro}$ is below $300$.
The lower  curves in Fig. 2a, identified by their ${\rm Ro}$ values,
delineate the region of rotational instability.
The largest growth rates are found for streamwise uniform
modes ($\alpha=0$) having a non-zero
spanwise wavenumber $\beta$ (Fig. 2b), that depends on the value of {\rm Ro},
in agreement with previous results for  other rotating  shear
flows. \cite{Hart,LezJon,AlfPer,YFMR}  The extent of the rotationally unstable region grows rapidly
with rotation, and is quite large at ${\rm Ro}=300$ which is the strongest rotation depicted in figure Fig. 2a.
Ultimately, this instability is suppressed for very strong rotation as examined below.
Note that the rotational instability takes on much larger growth rates than the TS instability, as illustrated by the
curves of constant growth rate shown in Fig. 2b for  the case ${\rm Re}=600$, ${\rm Ro}=300$.
More precisely the largest TS growth rate  is found in this case to be
 $\sup_{\alpha,\beta}(\omega_{i})_{TS}=0.006$
while the maximum value in Fig.2b is $\sup_{\alpha,\beta}\omega_{i}=0.145$

\smallskip

As in RB flow, an anti-cyclonic rotation for RAS leads, if not too strong, to a
rotationally-induced   instability which also reaches its maximum at  $\alpha=0$,
while a cyclonic rotation (${\rm Ro}<0$) is stabilizing.
These rotationally unstable regions in $(\alpha,\beta)$-plane are
presented in Fig.~3a for ${\rm Re}=600$ and  several  Rossby numbers ${\rm Ro}$.
Note that growth rates of the rotational instability for RAS are of comparable
magnitude to those found for RB flow (see Fig.3b for the case ${\rm Ro}=300$ where
$\sup_{\alpha,\beta}\omega_{i}=0.144$ is found).
On the contrary, no TS neutral curves appear for RAS (Fig.~3a) for ${\rm Re}=600$.
Recall that the AS profile is known to be much more stable
with respect to  TS instability compared with non-rotating Blasius flow.  This effect is
very apparent in the large critical Reynolds number of the AS profile (${\rm Re_c}=54370$); \cite{DR}
RAS flow thus lacks TS instability at ${\rm Re}=600$.

\smallskip
{\it Re-stabilization -- }

In RB flow, the condition for instability (\ref{vortcond})
corresponds to $1/{\rm Ro}< 0.5714$\cite{foot57} while in RAS flow the
condition becomes $1/{\rm Ro}<1$. The re-stabilization of rotating boundary layer flows by strong
anti-cyclonic rotation is confirmed by numerical computations. The curves in Fig.~4 represent
the largest growth rates ${\rm max}~\omega_i$ of this rotational instability found in the $(\alpha,\beta)$-plane
as a function of $1/{\rm Ro}$ (this  maximum  was always found to occur for $\alpha=0$).
At the largest Reynolds number depicted in Fig. 4a, ${\rm Re}=6000$, re-stabilization of RB flow
occurs at $1/{\rm Ro}\approx 0.57$ while for RAS, re-stabilization
occurs at $1/{\rm Ro}\approx 0.98$ when ${\rm Re}=60000$,
in agreement with the predicted values.
Large positive rotations have also been found to  re-stabilize channel flows,
\cite{LezJon,MatAlf2001} at ${\rm Ro}$ values consistent with
the vorticity criterion (\ref{vortcond}).

Since the rotational instability is fundamentally inviscid,
the minimum Reynolds number below which the rotational
instability is suppressed is quite low. The minimum
Reynolds numbers below which no instability is found
are ${\rm Re_C\approx 21}$ at ${\rm Ro_C\approx 10.9}$ for RB~(see Fig. 4a)
flow and ${\rm Re_C\approx 25.9}$ at ${\rm Ro_C\approx 23.5}$
for RAS flow~(see Fig. 4b).

\subsection{Transient Amplification}

The effect of rotation on transient growth for the RB and RAS flows is quantified through the analysis
of amplification gain $G_{MAX} (\alpha,\beta,{\rm Ro},{\rm Re})$
as a function of streamwise wavenumber $\alpha$, spanwise wavenumber $\beta$, Reynolds number and Rossby number.
Note that, in calculating $G_{MAX}$ and $G_P$ values, a
fixed time interval $[0,T]$ is used during which the maximum growth is
allowed to occur; the value of $T$ required to
capture the maximum varies with ${\rm Re}$ and is adjusted accordingly in
the calculations.

Non-rotating boundary layer flows are known to  display the following transient growth features:
(i) the most amplified disturbances tend to be
three-dimensional with $\beta\ne 0$ and $\alpha=0$;
(ii) the peak amplification for large Reynolds number  is proportional to the square of
the Reynolds number i.e.
\be
\label{scalings}
G_{MAX} \propto \frac{ {\beta}^2 {\rm Re}^2 T(k,\alpha {\rm Re})}{{k}^2} \; ,
\ee
where it is recalled that $k^2=\alpha^2+\beta^2$ and
$T$ is a function of $k$ and $\alpha {\rm Re}$ only;
(iii) the time to achieve peak amplification increases
linearly with the Reynolds number i.e. $t_{MAX}\propto {\rm Re}$.
For rotating boundary layers, scaling (\ref{scalings}) or its straightforward consequence
$G_P \propto {\rm Re}^2$  is  not preserved as  shown below.

For non-rotating boundary layers $(U(y),0,0)$, the Reynolds scalings
rely on the following statements (see also Ref.\onlinecite{SchHen}): upon performing the
rescalings ${\hat t} =t/{\rm Re}$, $ {\hat \zeta} =\zeta/(\beta {\rm Re})$, $ {\hat v} =v$,
equations (\ref{OStime})-(\ref{SQtime}) become dependent only on $k^2$ and $\alpha {\rm Re}$ and
the  energy turns into
\be
||{\bf q({\hat t})}||^2_E = \int_0^L (|D {\hat v}|^2 + k^2| {\hat v}|^2 ) dy  + (\beta {\rm Re})^2 \int_0^L ( | {\hat \zeta}|^2) dy \;.
\label{Enormmodif}
\ee
When the Rossby number is present, the energy expression is still valid but, in addition to $k^2=\alpha^2+\beta^2$
 and $\alpha {\rm Re}$, the parameters  ${\rm Ro}$ and
${(\beta {\rm Re})^2 \over{\rm Ro}}$ explicitly appear in  equations (\ref{OStime})-(\ref{SQtime})
\begin{widetext}
\begin{equation}
( \frac{\partial }{\partial {\hat t} } + i \alpha{\rm Re} U )(D^2-k^2){ {\hat v}} - i\alpha{\rm Re}  (D^2 U) { {\hat v}}
+ {1\over{\rm Ro}}i (\beta {\rm Re})^2 {\hat \zeta} - (D^2-k^2)^2{\hat v}=0 ,
\label{OStimemodif}
\end{equation}
\end{widetext}
\begin{equation}
(\frac{\partial }{\partial {\hat t}} + i\alpha {\rm Re} U  ){ \hat \zeta}
-{1\over{\rm Ro}}i{ \hat v} + i (DU){ \hat v}
- (D^2-k^2){ \hat \zeta}=0.
\label{SQtimemodif}
\end{equation}
The standard arguments are thus no longer valid when the Rossby number is present,
implying that the classical scalings will not be replicated.
The numerical computations which confirm this idea, are   presented below with
the observed scalings.

\smallskip
{\it Rotating Blasius boundary layer -- }

In Fig.5, level curves of $G_{MAX}$ in the $(\alpha,\beta)$-plane  are presented for RB flow in three cases:
weak cyclonic rotation ${\rm Ro}<0$
(Fig.5a), no rotation (Fig.5b), and weak anticyclonic rotation ${\rm Ro}>0$ (Fig.5c),
the Reynolds number being always fixed at ${\rm Re}=600$ which is
just above the critical value ${\rm Re}_c\approx 520$
for unstable Tollmien-Schlichting waves. The TS instability region is apparent
along the $\alpha$-axis centered at $\alpha \approx 0.3$ and
has only minor influence on the transient growth (see below).

In the absence of rotation, the peak
$G_{MAX}$ value was found to occur at $\alpha=0$ and $\beta=0.651$ (Fig.5b),
and to take  the value $G_P=545.8$ ( $G_P=545.9$ with $N=75$ polynomials and
$G_P=545.8$  with $N=125$). This value  is  in good agreement with other studies.
\cite{ButFar,BreKur,CorBot}  From previous computations by Butler and Farrell\cite{ButFar}
($G_P=1514$ at ${\rm Re}=1000$) or  Corbett and Bottaro\cite{CorBot} ($G_P=1515$ at ${\rm Re}=1000$)
the scaling law   $ G_P=1.515 \times 10^{-3} {\rm Re}^2 $ should be the correct one for non-rotating Blasius flow.
Using this latter formula, it is found that $G_P  = 545.4$ at ${\rm Re}=600$.

When the rotation is anti-cyclonic (Fig.5c) i.e. for ${\rm Ro}>0$,
rotational instability of \S IV(A) appears along the $\beta$-axis
and the level curves of $G_{MAX}$ are distorted in the neighborhood of
the highlighted  region of unstable normal modes. Moreover, the $G_{MAX}$ values
adjacent to the neutral curve are enhanced by the nearby modal instability,
and the peak amplification  occurs along the rotational neutral curve at
$\beta\approx 1$.  In such a case, transient growth
must compete with the strong modal instability.

With cyclonic rotation (Fig.5a) i.e. for ${\rm Ro}<0$, $G_{MAX}$ values are reduced in
magnitude everywhere in the $(\alpha,\beta)$-plane and the rotational instability is absent.
In addition, $G_P$ is now found at non-zero streamwise
wavenumber~(for ${\rm Ro}=-1800$ and ${\rm Re}=600$, $\alpha\approx 0.12$), while the spanwise wavenumber
slightly shifts from the value of the  non-rotating case~(for ${\rm Ro}=-1800$ and ${\rm Re}=600$,
$\beta\approx 0.75$). An optimal disturbance (hereafter denoted by OD) with weak streamwise
dependence ($\alpha\ne 0$) is often referred to as oblique, and is more typical of
plane Couette flow than of boundary layer flow.  Here, the obliquity is a
consequence of the Coriolis effects. The ${\rm Ro}$ dependence of the peak growth $G_P$
is illustrated in Fig.6. Each  point in this figure was obtained by first performing a
calculation of $G_{MAX}$ throughout the $(\alpha,\beta)$-plane (like
the one  of Fig.5a), then finding the peak value, $G_P$, and the
wavenumbers, $\alpha_P$ and $\beta_P$ at which the peak occurs.  Two
Reynolds numbers are shown: ${\rm Re}=600$, marked by open circles,
and ${\rm Re}=1800$, marked by dots.

The inset of Fig.6 shows the location in the $\alpha-\beta$ wavenumber
plane corresponding to the plotted $G_P$ values.  When rotation is very
weak, $G_P$ is found at $\alpha_P=0$, $\beta_P=0.651$; as rotation increases,
the location of $G_P$ loops through the plane, eventually settling
at $\beta_P=0$ for strong rotation, with a streamwise wavenumber $\alpha_P({\rm Re})$
dependent on  the Reynolds number.   The value  $G_P=16.70$
is found at $\alpha_P=0.467$ for ${\rm Re}=600$ while $G_P=51.02$
is found at $\alpha_P=0.364$ when ${\rm Re}=1800$.  These values are also recovered for
strong values of anti-cyclonic rotation.
The wavenumber $\alpha_P$ and $G_P$ values at strong rotation
agree with two-dimensional non-rotating transient growth
calculations.  Butler \& Farrell \cite{ButFar} give $G_{P}=28$ at $\alpha_P=0.42$
 for the best optimal 2D ($\beta_P=0$) disturbance  for non-rotating Blasius flow at
 ${\rm Re}=1000$. At the same Reynolds number, we find
$G_P=28.4$ for ${\rm Ro}=-0.5$
at $\alpha_P=0.42$.  These findings are consistent with the idea that
strong rotation imposes a two-dimensional Taylor-Proudman
r\'egime. \cite{LezJon,Mutabazi}

The ratio of the two $G_P$ curves in Fig.6 is a measure of the ${\rm Re}$
dependence of $G_P$ over a range of ${\rm Ro}$.  From the figure, this
dependence is clearly inconsistent with a $Re^2$ scaling throughout the entire
${\rm Ro}$ range. This ${\rm Re}$ dependence is explicitly shown in Fig.7a and
7c.  For weak rotation (${\rm Ro}=-10^6$), $G_P/ {\rm Re}^2 $ possesses a weak
${\rm Re}$ dependence~(Fig.7a) which strengthens for slightly stronger
rotations (${\rm Ro}=-10^5$).  In the strong rotation r\'egime~(Fig.7c), no
simple ${\rm Re}$ dependence is found for $G_P$ in RB flow: although $G_P$
seems linear in $ {\rm Re} $, a careful analysis reveals that this is not
strictly correct. In addition, figure Fig.7c confirms that, for transient
growth properties, the flow becomes increasingly two-dimensional in character
as rotation increases. Note that, in a similar manner, the time $t_p$ is not
linear in ${\rm Re}$ for weak rotation (${\rm Ro}=-10^6$)~(Fig.7b) while we
obtain scaling $t_P=0.78 {\rm Re}$ with no rotation. In the strong rotation
r\'egime~(Fig.7d), $t_P$ possesses no obvious scaling though one would guess
by eye a $\sqrt{{\rm Re}}$ law, which is not confirmed by a rigorous fit.

\smallskip
{\it Rotating asymptotic suction  flow -- }

The qualitative features found for RB flow are also recovered for RAS flow;
the following presentation can therefore be relatively terse.
Figure 8 is the equivalent of Fig. 5, but for RAS flow
at ${\rm Re}=600$. AS flow is well below its critical value $R_c=54370$
for TS instability, so no TS region is present. The most amplified
disturbances in the absence of rotation (Fig.8b) are again
streamwise uniform $(\alpha=0)$ while for weak cyclonic rotation
(Fig.8a) they begin to show weak streamwise variability
($\alpha\approx 0.07$ for the case of Fig.8a). Figure 9 is the RAS flow equivalent of Fig. 6; here, again, both ${\rm
Re}=600$, marked by open circles, and ${\rm Re}=1800$, marked by
dots, are calculated. These are the only known calculations of
transient growth in RAS flow. For AS flow (i.e. no-rotating case),  the values of
$G_P$ and $\beta_P$  have appeared in recent works:
Ref.~\onlinecite{FranssonCorbett} provides a scaling law of the form
$ G_P=0.99 Re^2$ which gives e.g. $ G_P=356.4$ at $Re=600$.
In the limit of very weak rotation, we find $G_P({\rm Re}=600)=358.12$ (or, when
the transverse velocity $V$ is included, $G_P({\rm Re}=600)=356.74$, in
good agreement with Ref.\onlinecite{FranssonCorbett}).
As in RB, scalings $G_P\propto {\rm Re}^2$ and $t_P\propto {\rm Re}$ are not recovered,
except for very weak rotation~(Fig.10a and 10b).
Again as in RB, a careful analysis  does not reveal  a clear scaling for $G_P$ and $t_P$
in  the strong  rotation r\'egime~(Fig.10c and 10d) though
 one would guess by eye   a scaling in  $ {\rm Re}  $ and $\sqrt{{\rm Re}}$ respectively.

The inset of Fig.9 shows the location in the $\alpha-\beta$ wavenumber
plane corresponding to the plotted $G_P$ values.  As in RB, the optimal
disturbance in RAS flow makes an excursion in wavenumbers, starting
at $\alpha_P=0$, $\beta_P=0.499$ ($\beta_P=0.530$ when $V$ is
included) for very weak rotation, looping through the plane and
settling on $\beta=0$ and a ${\rm Re}$-dependent $\alpha_P({\rm Re})$ when
rotation is strong. For instance, the value $G_P=5.736$ is found at $\alpha_P=0.670$
 for ${\rm Re}=600$ while  $G_P=13.12$ is found at $\alpha_P=0.617$
 for  ${\rm Re}=1800$. When $V$ is included, these values shift to
$G_P=5.661$ at $\alpha_P=0.675$  for
${\rm Re}=600$ and $G_P=13.01$ at $\alpha_P=0.621$  for
${\rm Re}=1800$.  These same values are also recovered for
strong values of anti-cyclonic rotation.

\subsection{Optimal Disturbances}

The most amplified disturbances in cyclonic RB and RAS flows
possess a non-null streamwise wavenumber: they thus correspond to
oblique vortices. The flow field of these disturbances can
be calculated from its expansion coefficients, already obtained in the
SVD solution of $G_{MAX}$. An example of the optimal
disturbance~(OD) in a weakly rotating RB flow, at ${\rm Re}=600$, ${\rm Ro}=-1\times 10^6$,
is depicted in Fig.11   and can be compared to the OD for the classical Blasius flow
at the same Reynolds.
In  figures Fig.11 and 12,  the velocity field  $v{\hat y}+w{\hat z}$ and the streamwise velocity
component $u$ are shown at a given streamwise location ($x=0$).
Due to the similarity of the optimal disturbances for RB
and RAS flows, RAS examples are not shown.

While the OD in RB at very weak rotation ${\rm Ro}=-1\times 10^8$ is
indistinguishable from the OD in non-rotating Blasius flow at both $t=0$ and
$t=t_P$, this is not so for larger but still weak rotations. Indeed
the OD in RB at ${\rm Ro}=-1\times 10^6$ (Fig.11 a) --which resembles
with apparent counter-rotating streamwise rolls, the OD in
non-rotating Blasius flow at $t=0$ (Fig.11c)-- 
is such that the extrema of streamwise velocity $u$ (Fig.11b) are
not precisely aligned with the up- and down-flows (Fig.11a) between the
vortices, as found in non-rotating Blasius flow (Figs.11c,d).  This offset
is still found in the OD at its greatest amplification at $t=t_P$.  Except
for a shift of phase, little change is seen in the $u$ field.
This explains why the $G_P$ value
is not too different from the one found in the non-rotating limit.
The kinetic energy of the OD is initially mainly in the $v$ or $w$
components and not in $u$ component.  At $t=t_P$ it is contained
mainly in $u$ field.

The above features are enhanced at stronger rotations as depicted in Fig.12,
 where ${\rm Ro}=-1800$: even the
OD at $t=0$ being profoundly modified. Finally, at very strong rotation (e.g. ${\rm Ro}=-1$  in Fig.13),
two-dimensionality is imposed  in the $x-y$ plane and the OD at $t=0$
(Fig.13) clearly exhibits the characteristic tilting against the
shear of the 2D Orr mechanism.  At $t=t_P$,  the shear has evolved
the initial disturbance into  vortices counter-rotating along the
$z$-axis (Fig.13).

\section{Conclusions}

This work has expanded on previous studies of boundary layer flows to
account for frame rotation. In particular, the temporal evolution of
three-dimensional disturbances  has been studied for
RB and RAS boundary layer flows in the presence of
spanwise rotation.  Normal mode and transient growth disturbances
properties were calculated and compared for the two flows.  Rotational
modal instability, similar to that known for rotating channel and free
shear flows, was found for weak anti-cyclonic rotation in RB and RAS;
neutral curves, growth rates and the critical point were calculated for
both flows.  Rotating boundary layer flows were also found to exhibit
transient growth. In the modally stable cyclonic regime the transient
growth properties strongly depend on the rotation rate. In particular:
(i) peak growth factors are reduced by rotation;
(ii) even at weak rotation, the classical $Re^2$
dependence is not recovered;
(iii) optimal disturbances take the form of oblique vortices which 
transform continuously from streamwise uniformity
($\alpha=0$) when rotation is very weak to spanwise uniformity ($\beta=0$)
when rotation is strong, this last feature being a consequence  of a Taylor-Proudman r\'egime.

It is anticipated that the results presented here may be useful in
dealing with geophysical and astrophysical flows, where rotation is
important and curvature is often negligible. Most noteworthy is the
example of Keplerian flow in an accretion disk, where no purely
hydrodynamic inviscid linear eigenvalue instability has been
identified. A study of transient growth in two-dimensional accretion disks
\cite{IK} has recently been performed and further inquiry is
already under way.  The problems analyzed in this work constitute an interesting
element of disk flow studies.

Finally let us mention that
the longitudinal roll instability in rotating shear flow exhibits
weak downstream development \cite{AlfPer}. Because
the development is weak and because a great deal of closely related previous work
\cite{Johnson,PotCha,YFMR,LezJon,Hart} 
is based on temporal theory, this study  adopted a temporal framework.
However recent spatial analyses of ordinary boundary layer
transient growth \cite{Luchini12,AndBerHen,TumRes}
are shown to be more realistic with respect to observations. \cite{White}
In future investigation, this aspect should be envisaged to complement
the present temporal theory.

\vskip 1cm
\noindent

\begin{acknowledgments}
Thanks to Ed Spiegel for enlightening discussions
and to the Aspen Center for Physics for its
hospitality and support at the inception of this work.
\end{acknowledgments}

\bigskip
\appendix{\bf APPENDIX: The transverse velocity in the presence of rotation.}

Transverse velocity $V$ is normally neglected in the classical
temporal stability analysis of Blasius flow, being of the same order as
other neglected terms.   The effect  of this assumption on the instability normal  modes
is well documented for non-rotating  AS  and Blasius  flows (see standard
references, e.g. Ref.~\onlinecite{DR}).
In this appendix,  we examine the effect of this assumption on two aspects which have not been
previously studied:
(a) the effect of $V$ on transient growth properties with or without rotation; (b) the effect of $V$ on the
rotational normal mode instability properties.   We consider  these two effects
on  the  AS and RAS flows since they are true Navier-Stokes solution when $V$ is present which authorizes a quantitative
comparison between the simplified analysis ($V=0$) and the complete one.

An overview of the effect of $V$ on the transient growth properties
of the   AS flow (i.e. non-rotating) can be found in Table I.  Briefly, the neglect of
$V$ has little effect on the peak growth factors $G_P$ of AS flow,
less than one percent.  The wavenumber of peak growth,
$\beta_P$, is shifted to larger values when $V$ is included in
AS calculations, but the shift is relatively small.  The two
cases are illustrated in Fig. 14 where level surfaces of $G_P(\alpha,\beta)$
are shown for the  AS calculated with and without the $V$ term.
As regard  $t_P$ values,  we   obtain $t_P=1.145 {\rm Re} $ when V is included 
which agrees with Ref.\onlinecite{FranssonCorbett}
and we  find  $t_P=1.331 {\rm Re}$   when V is neglected.

The effect of $V$ on the rotational instability neutral curves and
growth rates  is small, except near the critical point $({\rm Re}_C, {\rm Ro}_C)$.
These properties are illustrated for RAS in Fig.15a, showing the shifted
neutral curves, and Fig.15b, showing the modified growth rates.
When $V$ is included, the location of the critical point shifts
from ${\rm Re}_C=25.9, {\rm Ro}_C=23.5$ to ${\rm Re}_C=33.7, {\rm Ro}_C=14.6$,
as seen in Fig.16 (compare to Fig.4(b)).

Under rotation, the effect of $V$ on the transient growth properties
of the RAS remains negligible.  To show this we display in Fig.17 the
differences in peak growth factors $G_P$ calculated with and without
the $V$ term for ${\rm Re}=600, {\rm Ro}=1800$.

\clearpage
\newpage

\begin{table}
\caption{\label{tab:table2}
Peak transient growth quantities for  non-rotating AS flows
computed with and without the transverse velocity, $V$ term.  Values for non-rotating  Blasius flow with
$V=0$ are given for the sake of completeness.
}
\begin{ruledtabular}
\begin{tabular}{ccccc}
& Blasius & Blasius & AS & AS \\
& $Re=600$ & $Re=1800$ & $Re=600$ & $Re=1800$ \\
\hline
$G_P(V=0)$& 545.93 & 4910.5 & 358.12 & 3220.2 \\ 
$G_P(V\ne 0)$& --- & --- & 356.74 & 3207.5 \\ 
\hline
$\beta_P$(V=0)& 0.651 & 0.651 & 0.499 & 0.499\\
$\beta_P(V\ne 0)$& --- & --- & 0.529 & 0.530\\
\end{tabular}
\end{ruledtabular}
\end{table}
\centerline{Table I, Yecko, Phys. Fluids}

\clearpage
\newpage
\centerline{\bf List of Figure Captions}

\bigskip

\noindent
{FIG. 1: Sketch of the flow configuration.}

\smallskip

\noindent
{FIG. 2:
(a) Neutral curves  of RB flow at Reynolds number ${\rm Re}=600$ and
various Rossby numbers ${\rm Ro}$; $\alpha$ and $\beta$ are the streamwise and spanwise wavenumber, respectively.
(b) Curves of constant growth rate  $\omega_i$ for RB flow at ${\rm Re}=600$ and ${\rm Ro}=300$.
}

\smallskip

\noindent
{FIG. 3:
(a) Neutral curves of RAS flow at Reynolds number ${\rm Re}=600$ and various $Ro$;
(b) Curves of constant growth rate  $\omega_i$ for RAS flow at ${\rm Re}=600$ and ${\rm Ro}=300$.
}

\smallskip

\noindent
{FIG. 4:
(a) Largest growth rates ${\rm max}~\omega_i$
for anti-cyclonic RB flow at Reynolds numbers
$Re=25,60,600,6000$ (curves bottom to top); Inset: projection of the
neutral surface $\omega_i(\beta,Re,Ro)=0$ onto the $Re-Ro$ plane, showing
the critical point $Re_C=21.005, Ro_C=10.9$;
(b) Same as (a) but for RAS at $Re=40,60,600,6000,60000$ (curves bottom to top);
Inset: critical point $Re_C=25.9, Ro_C=23.5$.
}

\smallskip

\noindent
{FIG. 5:
Level curves of $G_{MAX}$ in the $(\alpha,\beta)$-plane for RB flow at ${\rm Re}=600$ and  for weak
cyclonic $({\rm Ro}=-1800)$ rotation (a), no rotation (b), and weak
anti-cyclonic $({\rm Ro}=1800)$ rotation (c). Regions of modal instability are shaded.
}

\smallskip

\noindent
{FIG. 6:
Peak  amplification factors $G_P$ of RB flow at ${\rm Re}=600$ and ${\rm Re}=1800$ as a function of
cyclonic ${\rm Ro}<0$; Inset: corresponding location in the wavenumber $\alpha-\beta$ plane of
the optimal disturbance associated with $G_P$ values in the figure; weak rotation
cases lie along the $\beta$ axis while strong rotation cases lie along $\alpha$.
}

\smallskip

\noindent
{FIG. 7: (a)  $G_P/{\rm Re}^2$  of RB flow as a function of
${\rm Re}$ for two weak rotation cases ${\rm Ro}=-10^5$, ${\rm Ro}= -10^6$ and no-rotation;
(b) $t_P/Re$  of RB flow as a function of ${\rm Re}$ for the same cases as (a);
(c) $G_P$ of RB flow as a function of
${\rm Re}$ for strong rotation, ${\rm Ro}=-1$, and two-dimensional disturbances;
(d)  $t_P$ of RB flow as a function of ${\rm Re}$ for the same cases as (c).
}

\smallskip

\noindent
{FIG. 8:
Level curves of $G_{MAX}$ in the $(\alpha,\beta)$-plane for  RAS flow at ${\rm Re}=600$ for weak
cyclonic $({\rm Ro}=-1800)$ rotation (a), no rotation (b), and weak
anti-cyclonic $({\rm Ro}=1800)$ rotation (c). Regions of modal instability are shaded.
}

\smallskip

\noindent
{FIG. 9: Peak amplification factors $G_P$ of RAS flow at ${\rm
Re}=600$ and ${\rm Re}=1800$ as a function of cyclonic ${\rm Ro}<0$;
Inset: corresponding location in the $\alpha-\beta$ plane of the
optimal disturbance associated with $G_P$ values in the figure; weak
rotation cases lie along the $\beta$ axis while strong rotation cases
lie along the $\alpha$ axis.  }

\smallskip

\noindent
{FIG. 10: (a)  $G_P/{\rm Re}^2$  of RAS flow as a function of
${\rm Re}$ for two weak rotation cases ${\rm Ro}=-10^5$, ${\rm Ro}= -10^6$ and no-rotation;
(b) $t_P/Re$  of RAS flow as a function of ${\rm Re}$ for the same cases as (a);
(c) $G_P$ of RAS flow as a function of
${\rm Re}$ for strong rotation, ${\rm Ro}=-1$, and two-dimensional disturbances;
(d)  $t_P$ of RAS flow as a function of ${\rm Re}$ for the same cases as (c).
}

\smallskip

\noindent
{FIG. 11: (a) Velocity vectors $v{\hat y}+w{\hat z}$
of the most amplified disturbance in cyclonic RB
flow at ${\rm Re}=600$ and ${\rm Ro}=-10^6$.
The disturbance is shown at $t=0$ (left) and at
the time of maximum growth (right).
(b) Streamwise velocity $u$ for OD of (a).
(c) Velocity vectors $v{\hat y}+w{\hat z}$ for
non-rotating Blasius flow at ${\rm Re}=600$.
The disturbance is shown at $t=0$ (left) and at
the time of maximum growth (right).
(d) Streamwise velocity $u$ for OD of (c).
}

\smallskip

\noindent
{FIG. 12:
(a) Velocity vectors $v{\hat y}+w{\hat z}$
of the most amplified disturbance in cyclonic RB
flow at ${\rm Re}=600$ and ${\rm Ro}=-1800$.
The disturbance is shown at $t=0$ (left) and at
the time of maximum growth (right).
(b)Streamwise velocity $u$ for OD of (a).
}

\smallskip

\noindent
{FIG. 13:
Velocity vectors $u{\hat x}+v{\hat y}$ in the $x-y$ plane
of the most amplified disturbance in cyclonic RB
flow at ${\rm Re}=600$ and ${\rm Ro}=-1$.
The disturbance is shown at $t=0$ (left) and at
the time of maximum growth (right).
}

\smallskip

\noindent
{FIG. 14:
Comparison of the level curves of $G_{MAX}$ in the $(\alpha,\beta)$-plane
for AS flow calculated with $V$ included (solid curves) and $V$ neglected (dashed).
The peak growth is found at $\alpha_P=0$ and $\beta_P=0.529$ (V included) or $\beta_P=0.499$ (V neglected).
}

\smallskip

\noindent
{FIG. 15:
(a) Neutral curves of RAS flow at Reynolds number ${\rm Re}=600$ and three $Ro$;
(b) Curves of constant growth rate  $\omega_i$ for RAS flow at ${\rm Re}=600$ and ${\rm Ro}=300$.
In the figure, the dashed curves correspond to calculations
in which $V$ was neglected, reproducing Fig.3; the solid
curves were computed with $V$ included.
}

\smallskip

\noindent
{FIG. 16:
Largest growth rates ${\rm max}~\omega_i$
for anti-cyclonic RAS flow with $V$ included at Reynolds numbers
$Re=40,60,600,6000,60000$ (curves, bottom to top); Inset: projection of the
neutral surface $\omega_i(\beta,Re,Ro)=0$ onto the $Re-Ro$ plane, showing
the critical point at ${\rm Re}=33.77$, ${\rm Ro}=14.5$. To be compared to Fig.4(b).
}

\smallskip

\noindent
{FIG. 17: Comparison of maximum transient growth level curves for
RAS flow at ${\rm Re} =600, {\rm Ro} =1800$,
calculated with $V$ included (solid curves) and $V$ neglected (dashed curves).
The peak growth factor is $G_P=130.79$ ($V$ included) or
$G_P=129.86$ ($V$ neglected).
}

\clearpage
\newpage
\begin{figure*}
\begin{center}
  \includegraphics{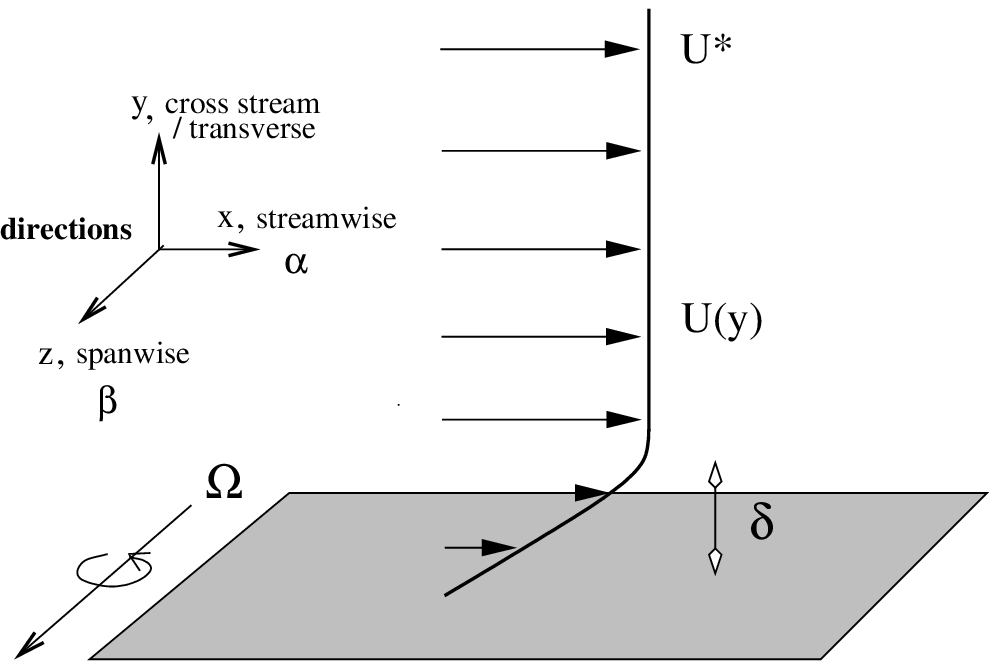}
\end{center}
\caption{Yecko, Phys. Fluids}
\label{schemat}
\end{figure*}


\clearpage
\newpage
\begin{figure*}
\begin{center}
  \includegraphics[width=3.000in]{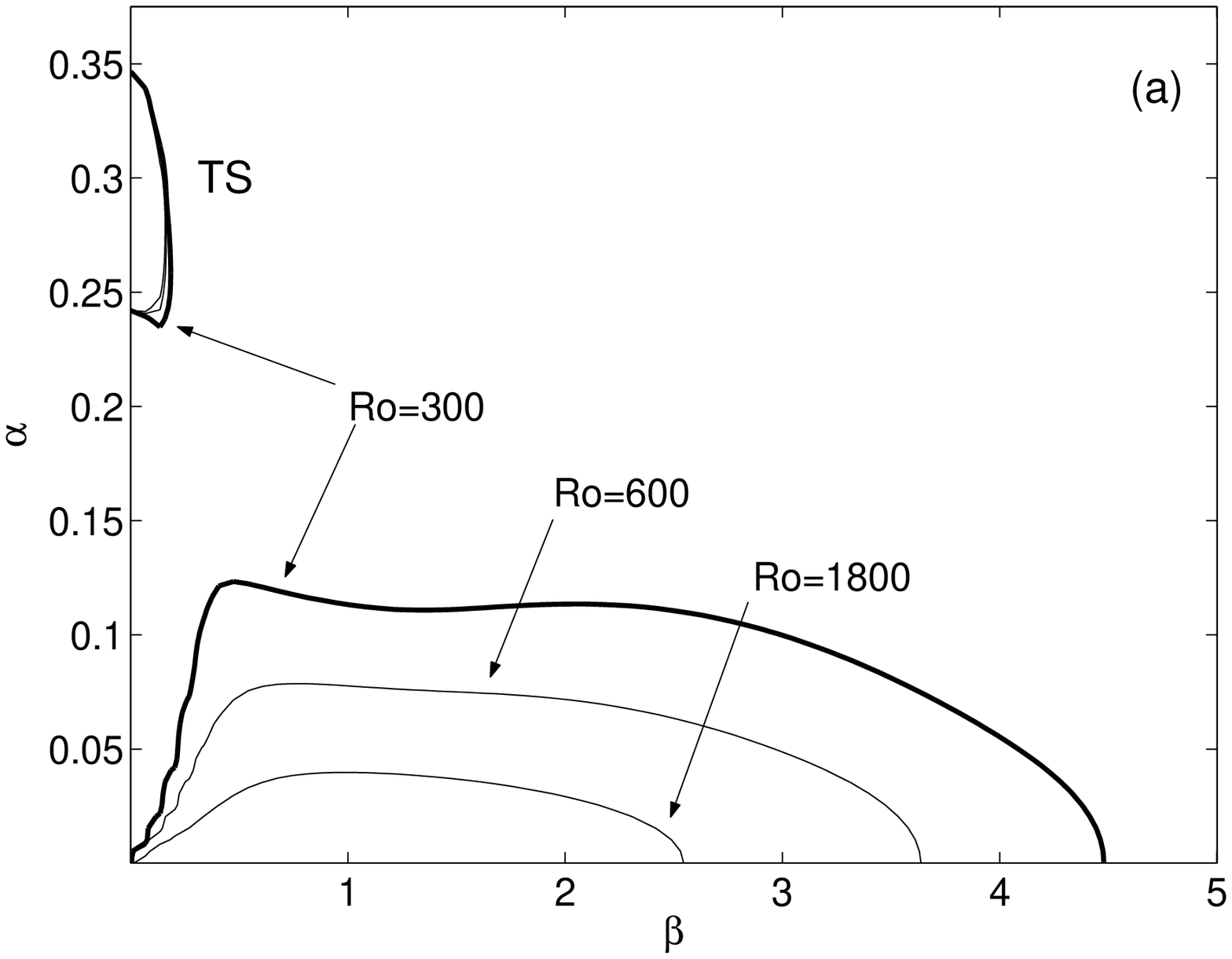}
  \includegraphics[width=3.000in]{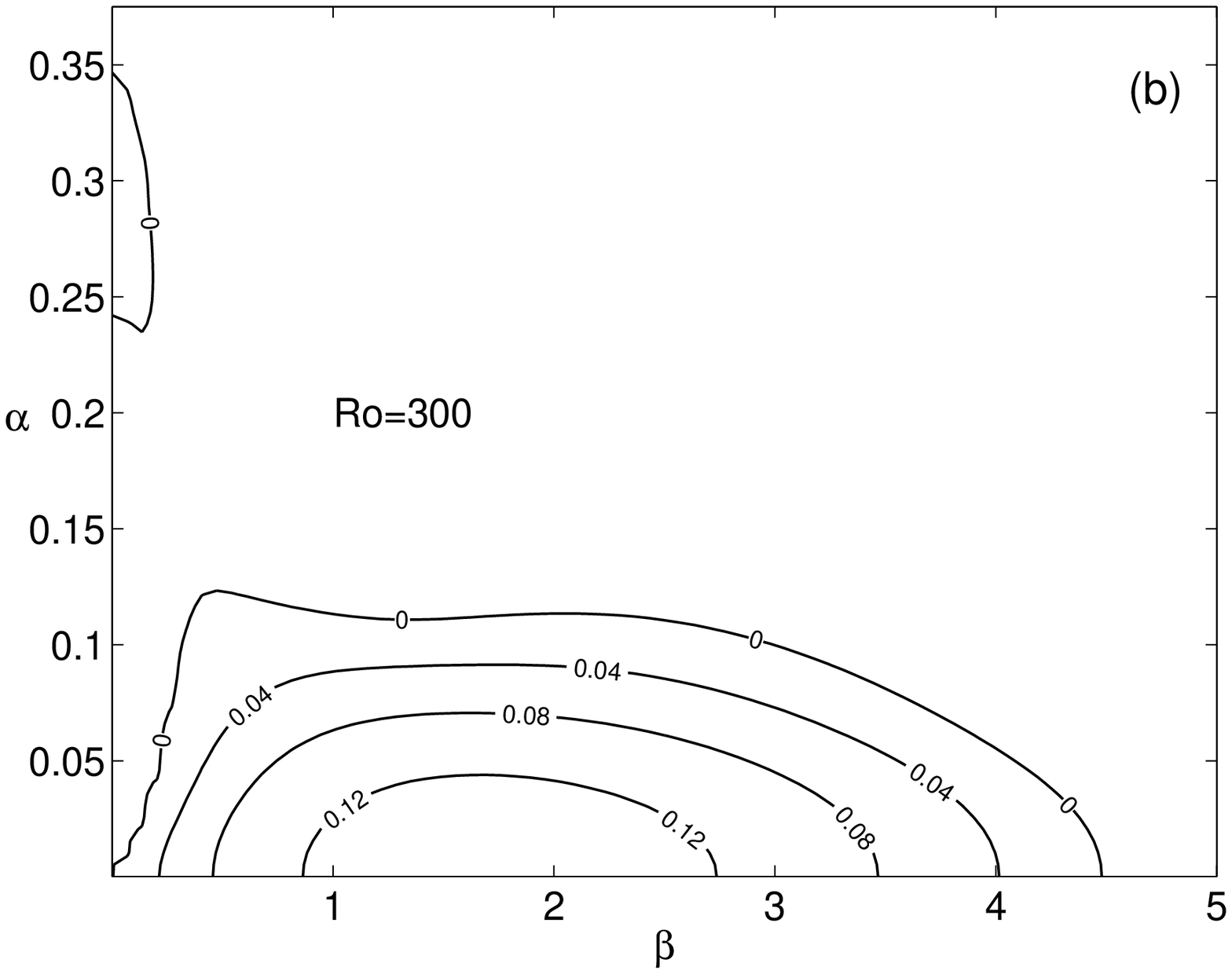}
\end{center}
\caption{Yecko, Phys. Fluids}
\label{blaevfig}
\end{figure*}

\clearpage
\newpage
\begin{figure*}
\begin{center}
  \includegraphics[width=3.000in]{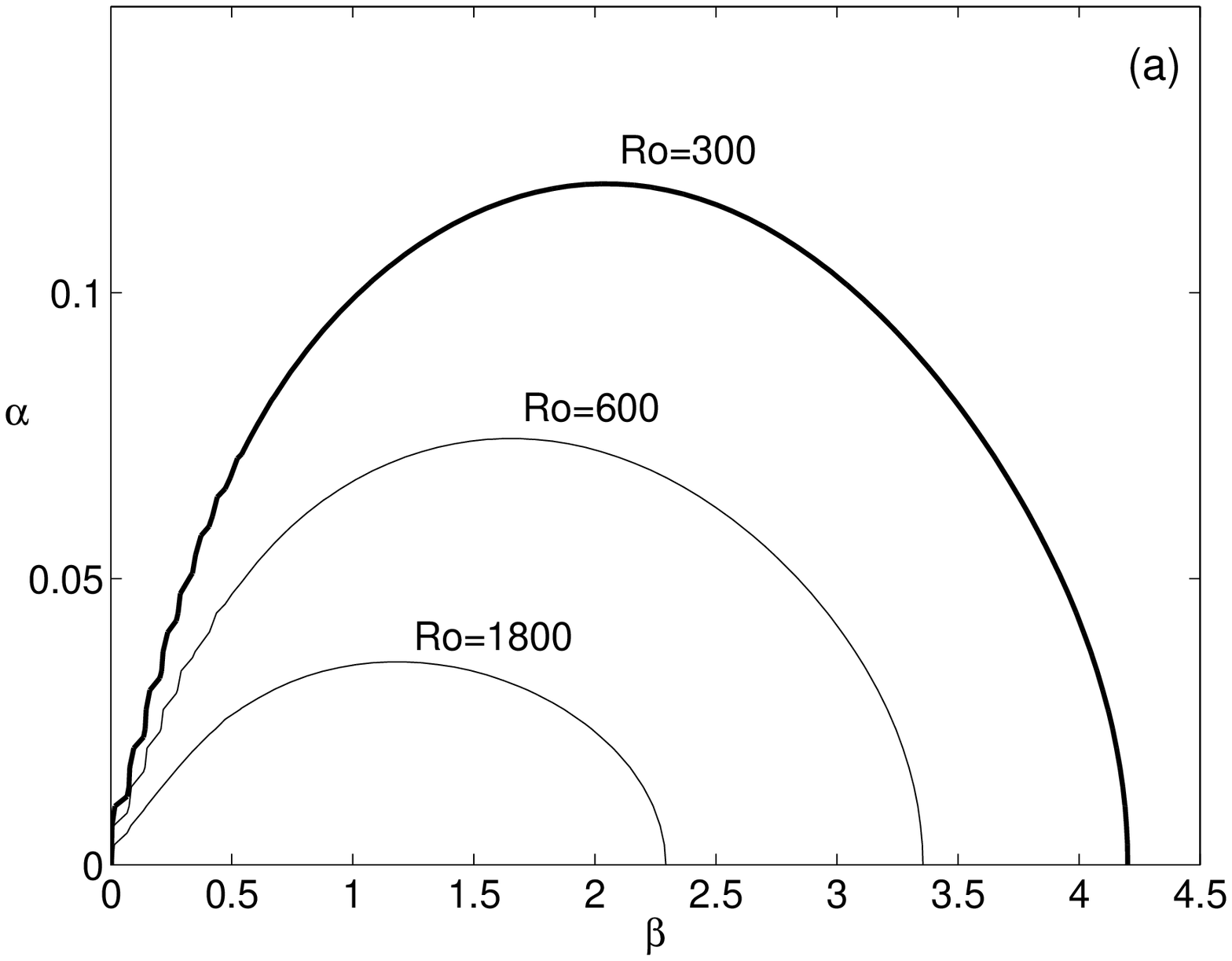}
  \includegraphics[width=3.000in]{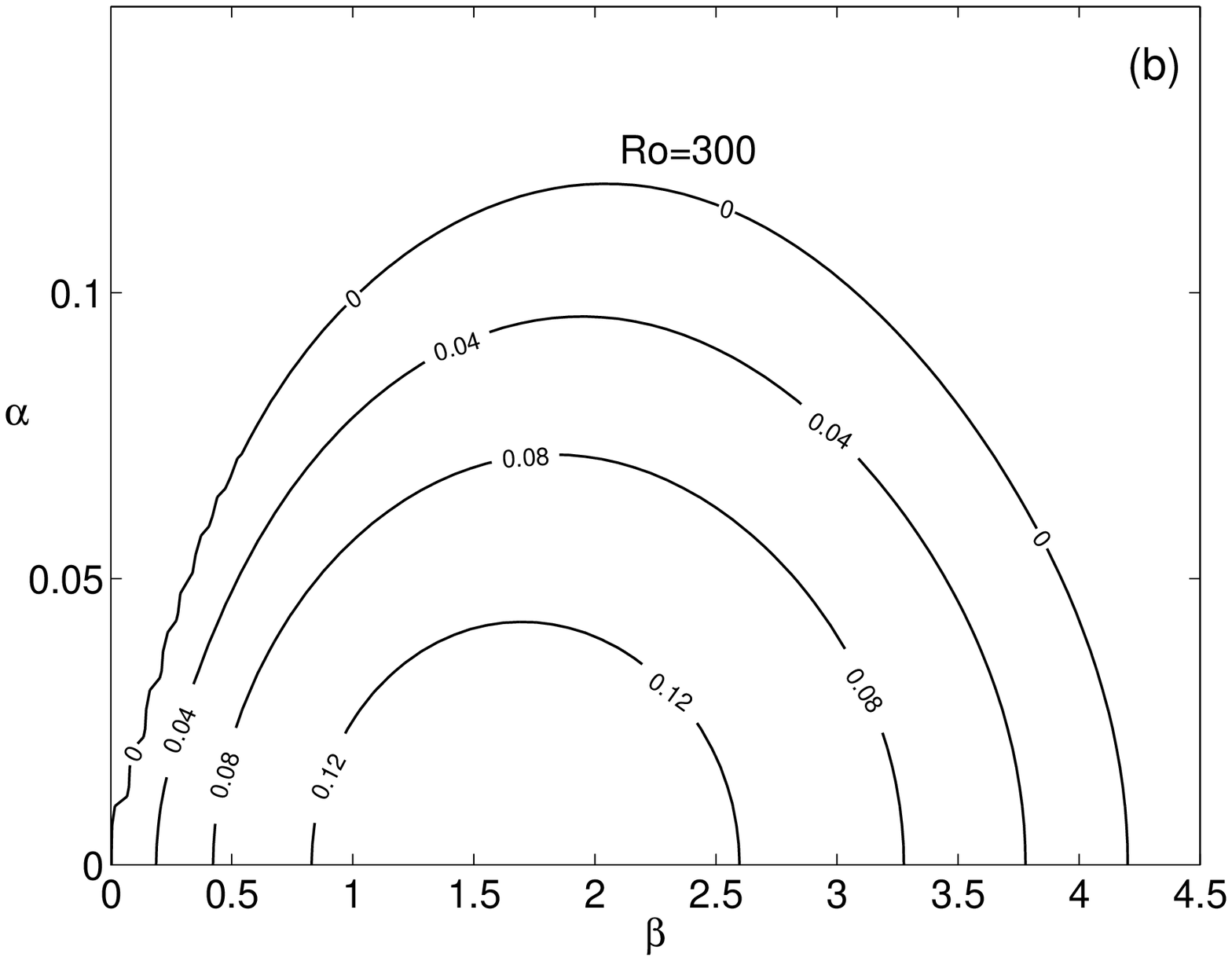}
\end{center}
\caption{Yecko, Phys. Fluids}
\label{suxevfig}
\end{figure*}

\clearpage
\newpage
\begin{figure*}
\begin{center}
  \includegraphics[width=3.000in]{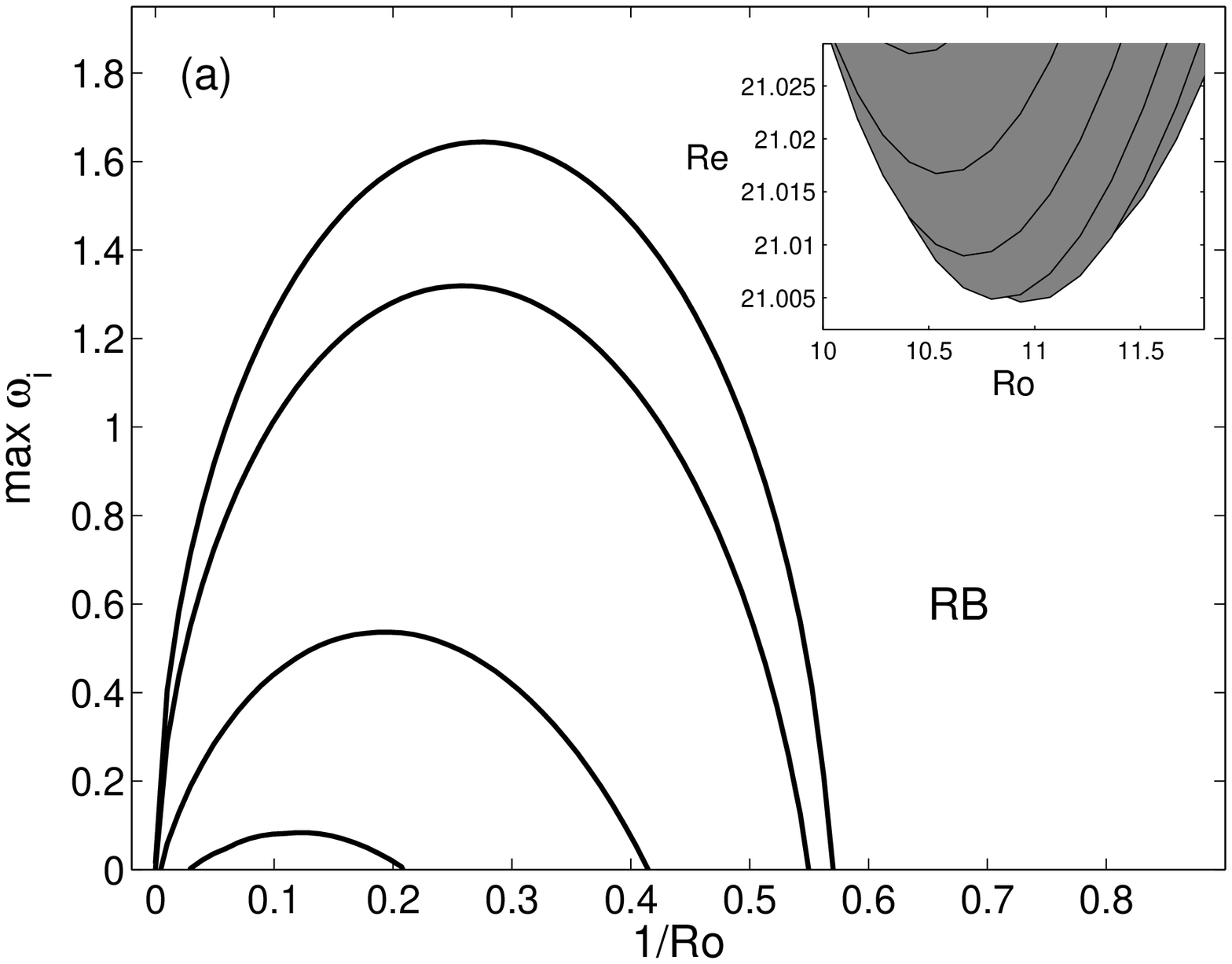}
  \includegraphics[width=3.000in]{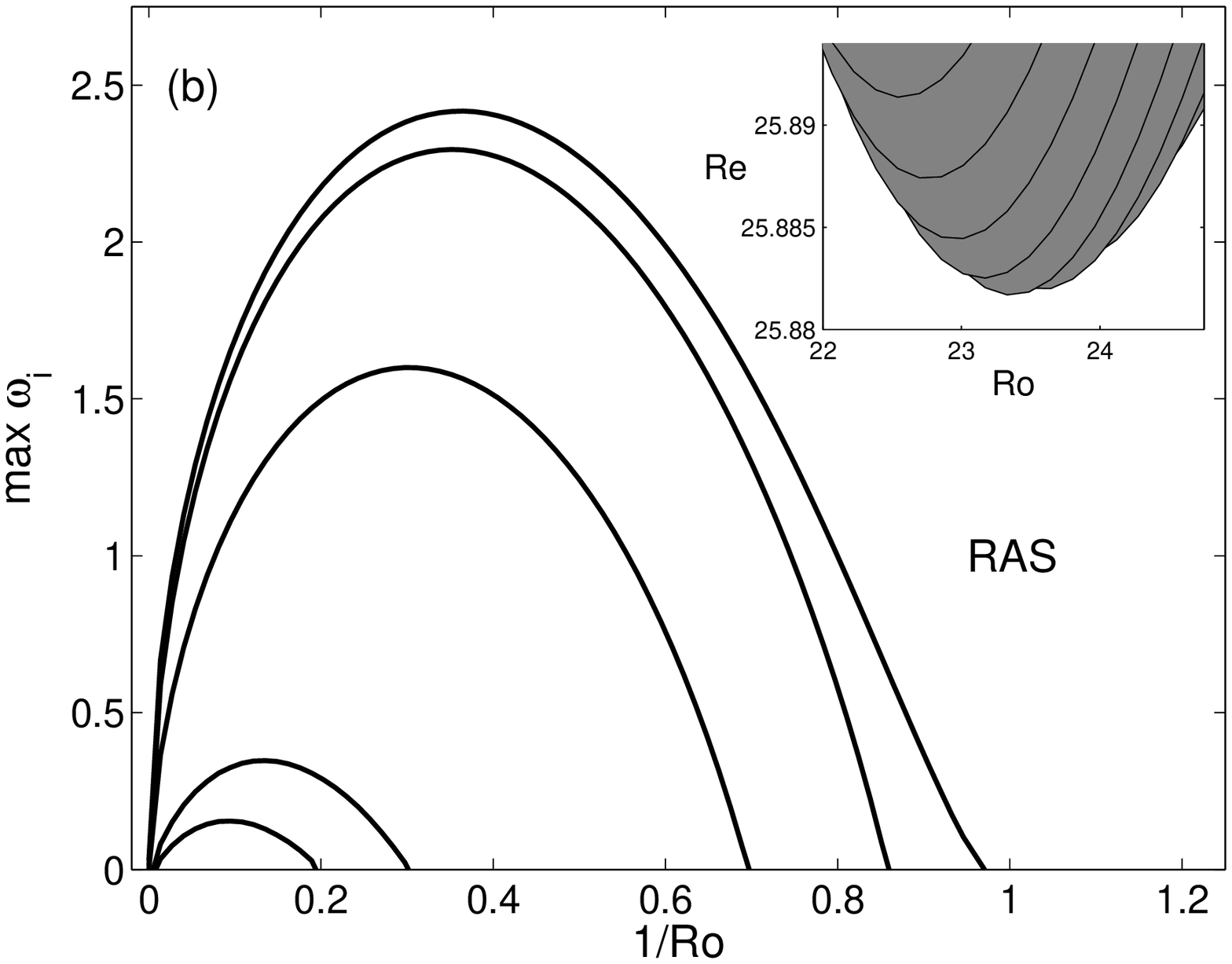}
\end{center}
\caption{Yecko, Phys. Fluids}
\label{blasRofig}
\label{suxRofig}
\end{figure*}

\clearpage
\newpage
\begin{figure*}
\begin{center}
  \includegraphics[width=3.0in]{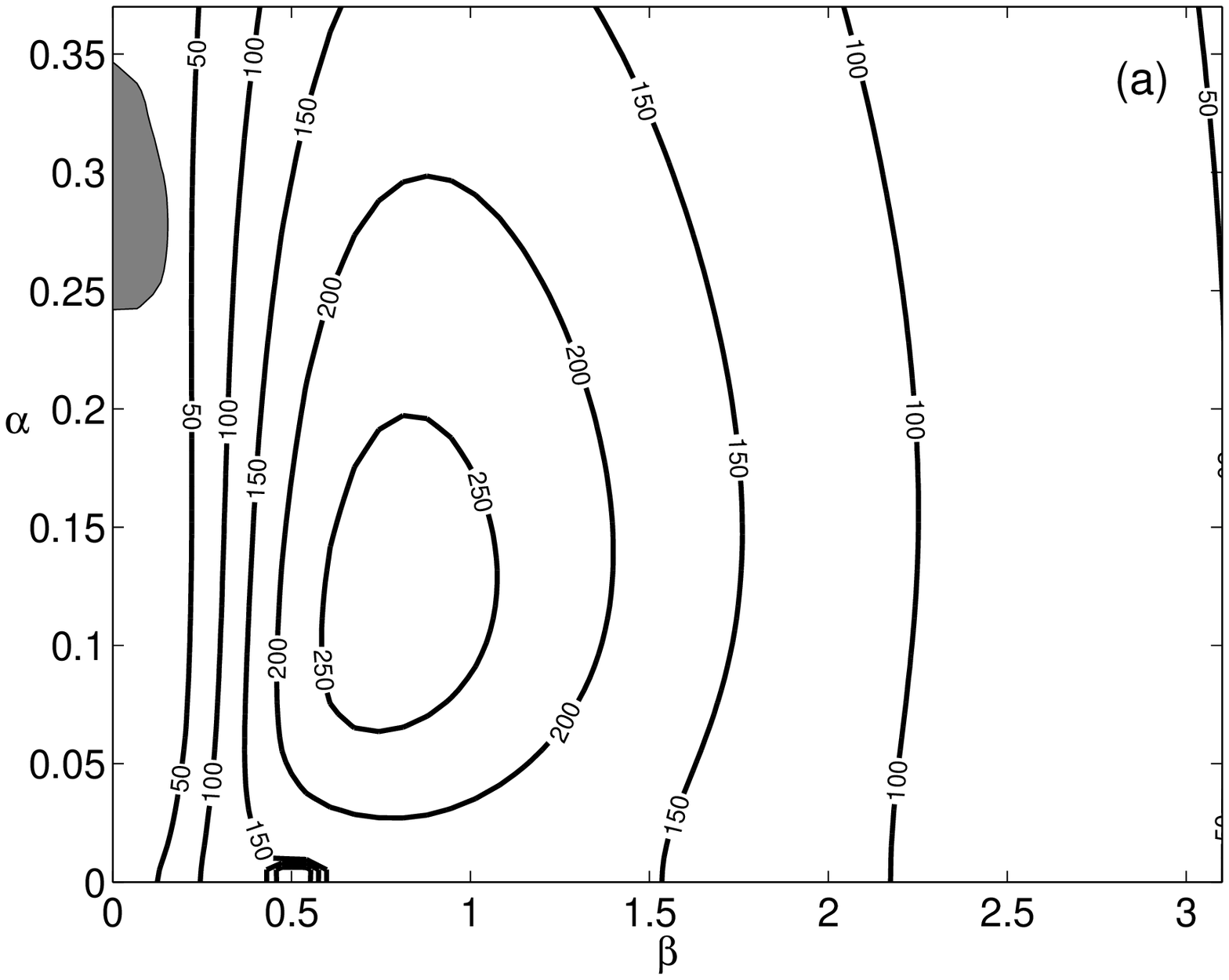}

  \includegraphics[width=3.0in]{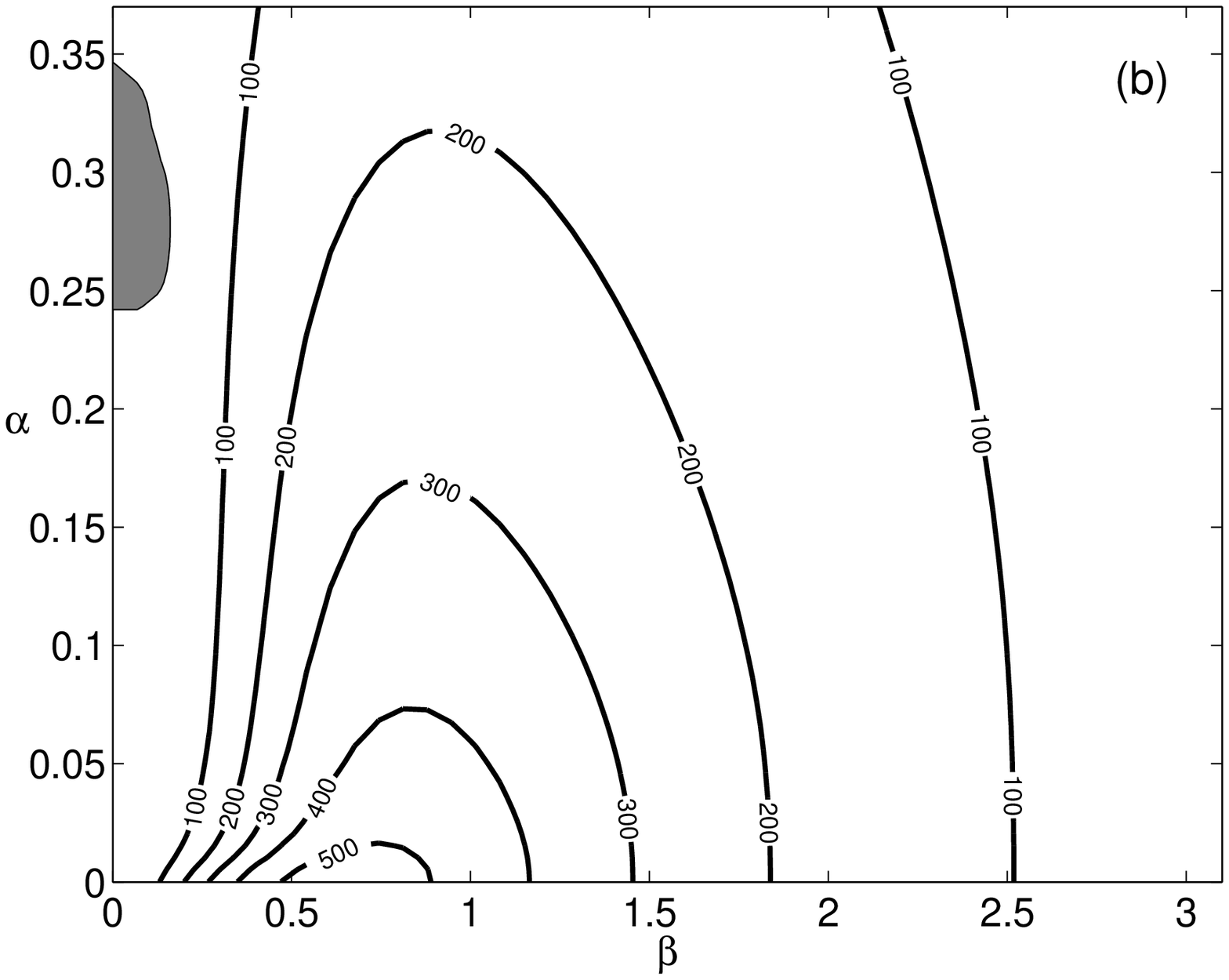}

  \includegraphics[width=3.0in]{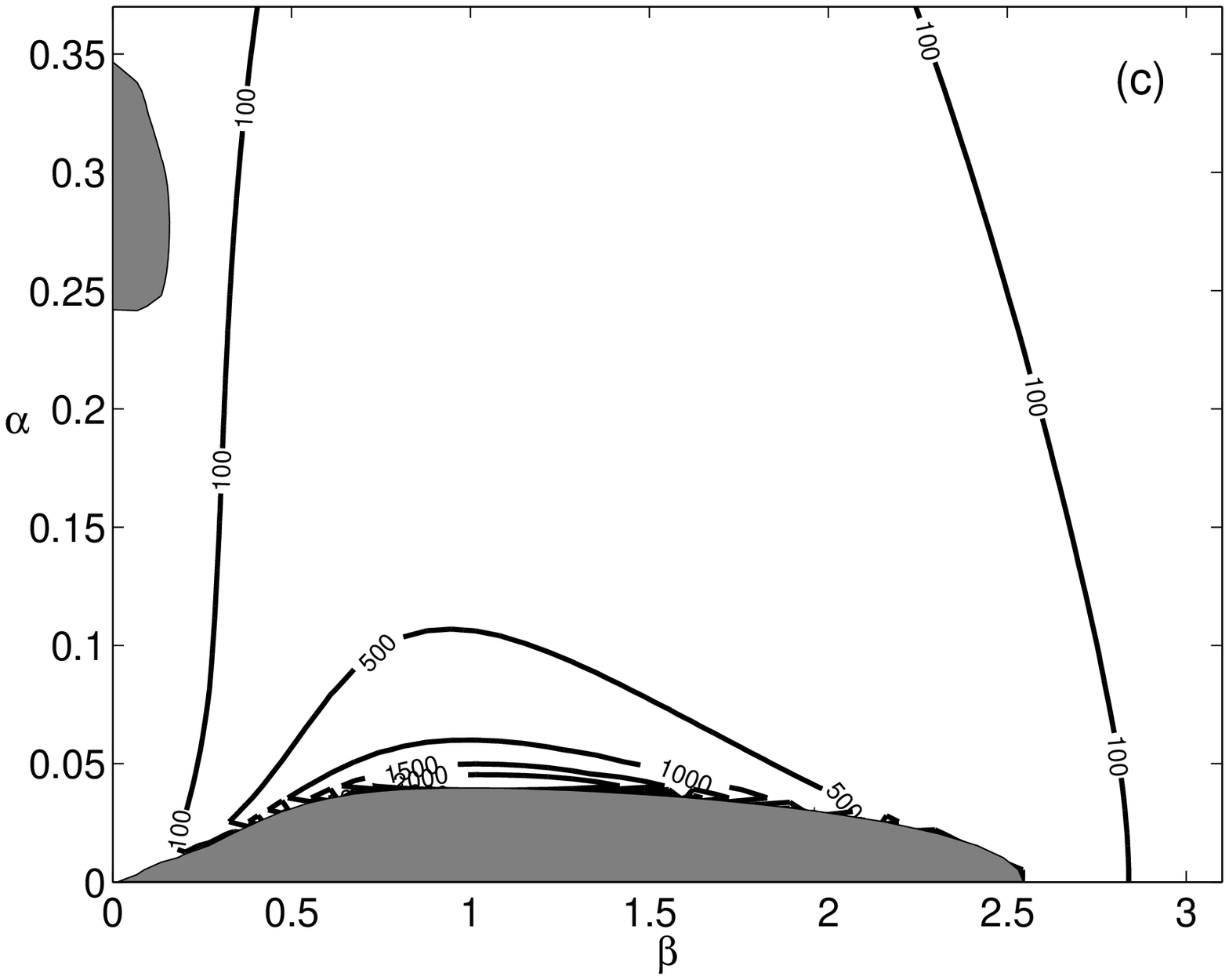}
\end{center}
\caption{Yecko, Phys. Fluids}
\label{roblasiusfig}
\end{figure*}

\clearpage
\newpage
\begin{figure*}
\begin{center}
  \includegraphics[width=3.25in]{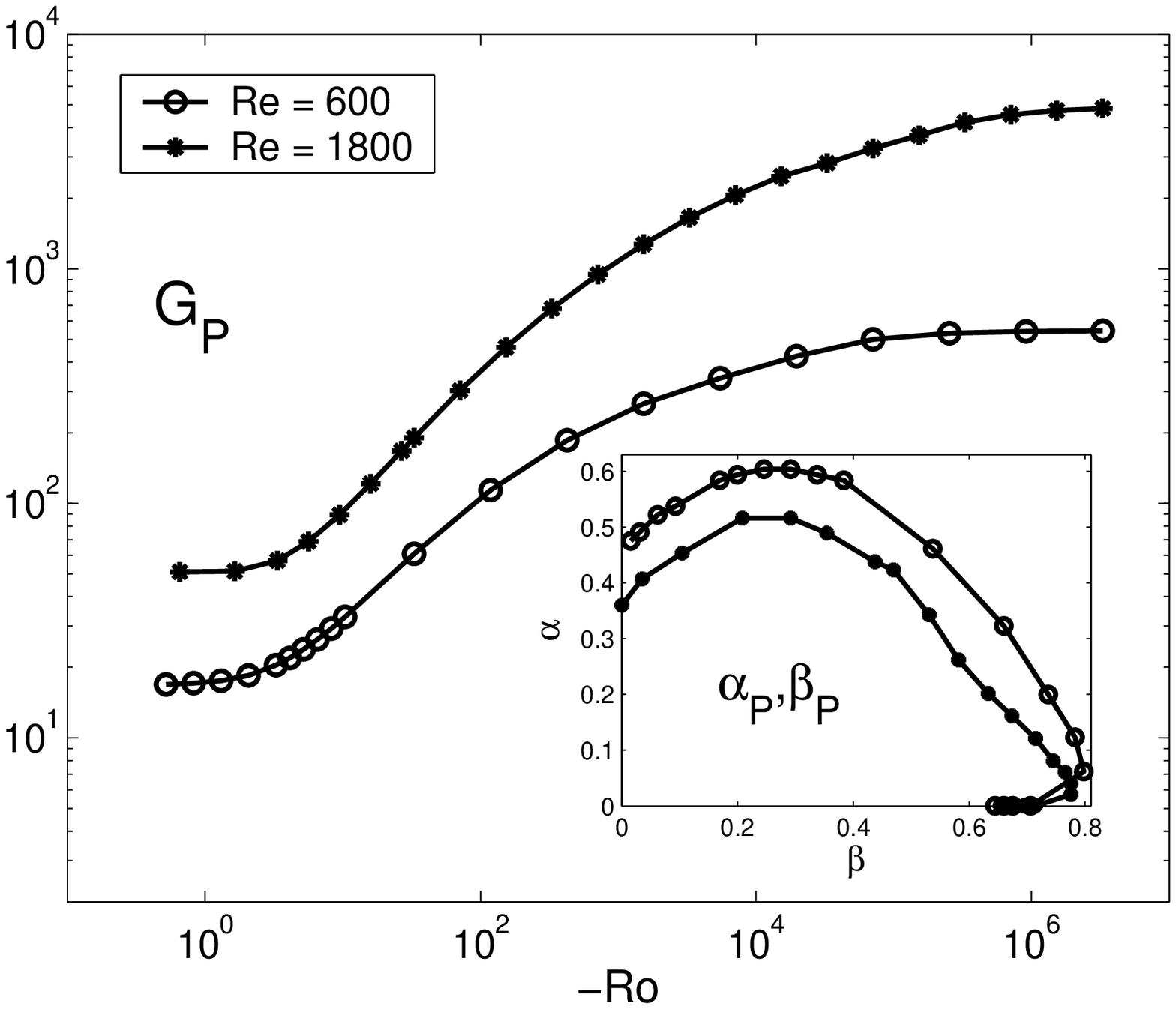}
\end{center}
\caption{Yecko, Phys. Fluids}
\end{figure*}

\clearpage
\newpage
\begin{figure*}
\begin{center}
  \includegraphics[width=3.000in]{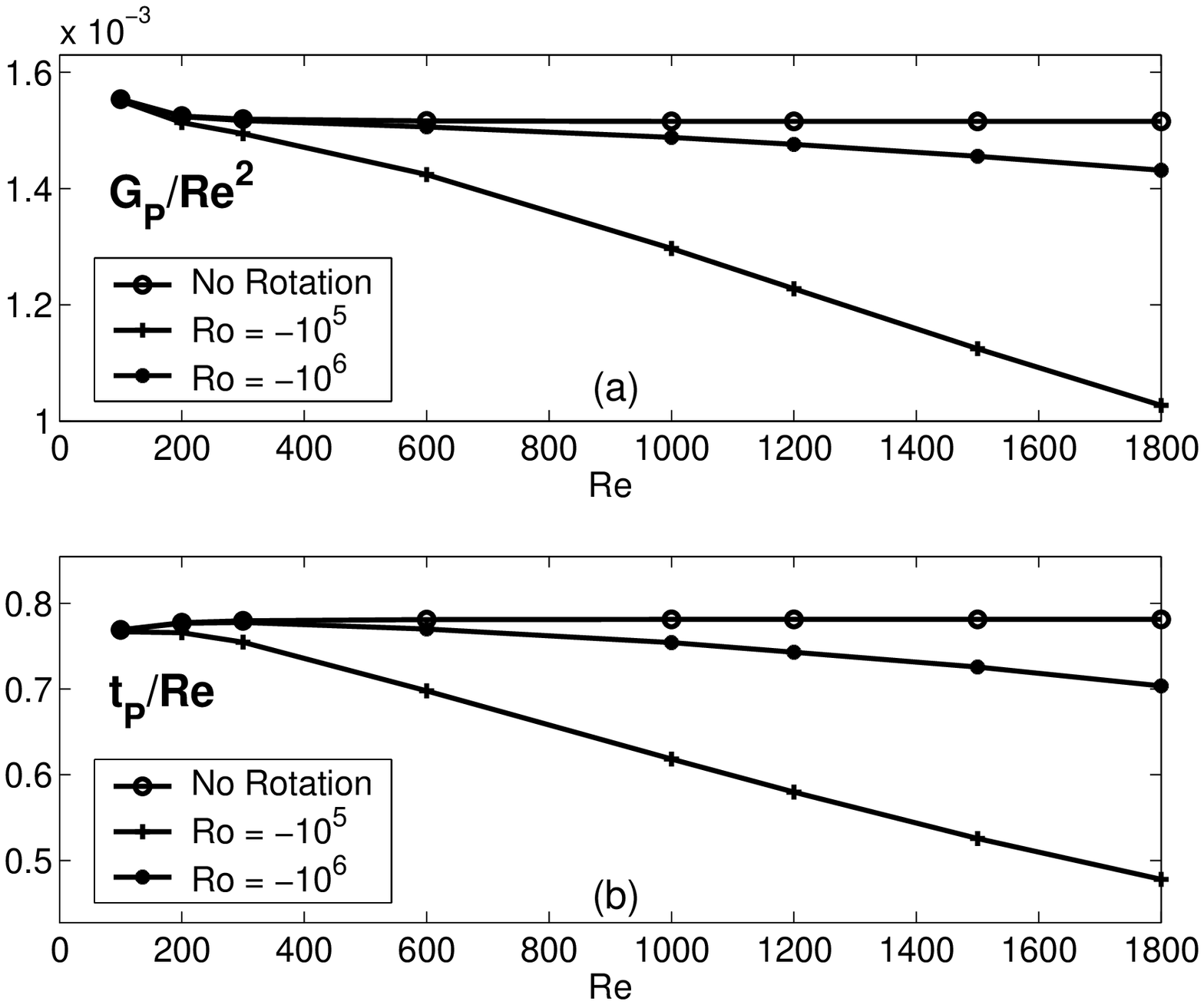}
  \includegraphics[width=3.000in]{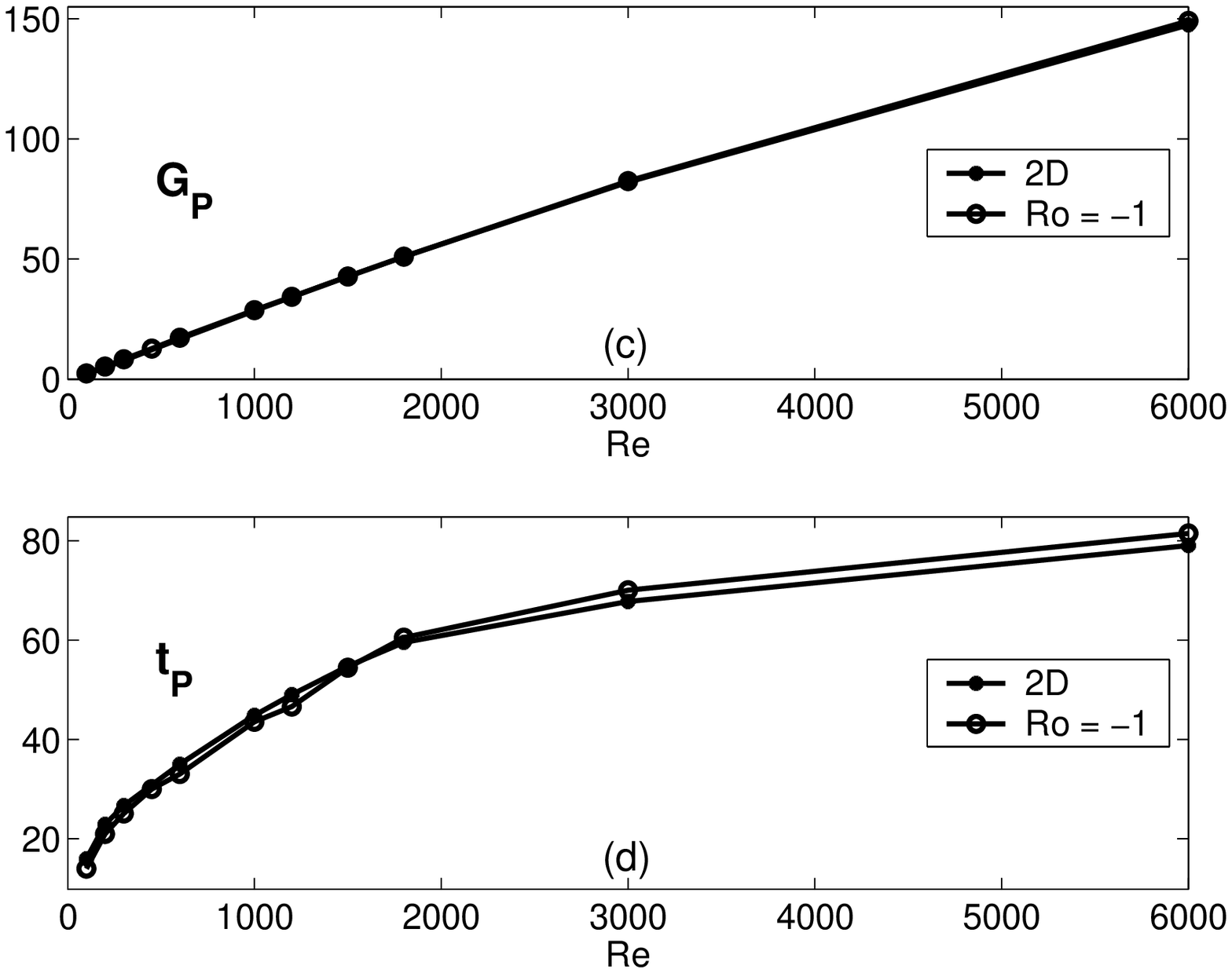}
\end{center}
\caption{Yecko, Phys. Fluids}
\label{fig7BIS}
\end{figure*}

\clearpage
\newpage
\begin{figure*}
\begin{center}
  \includegraphics[width=3.0in]{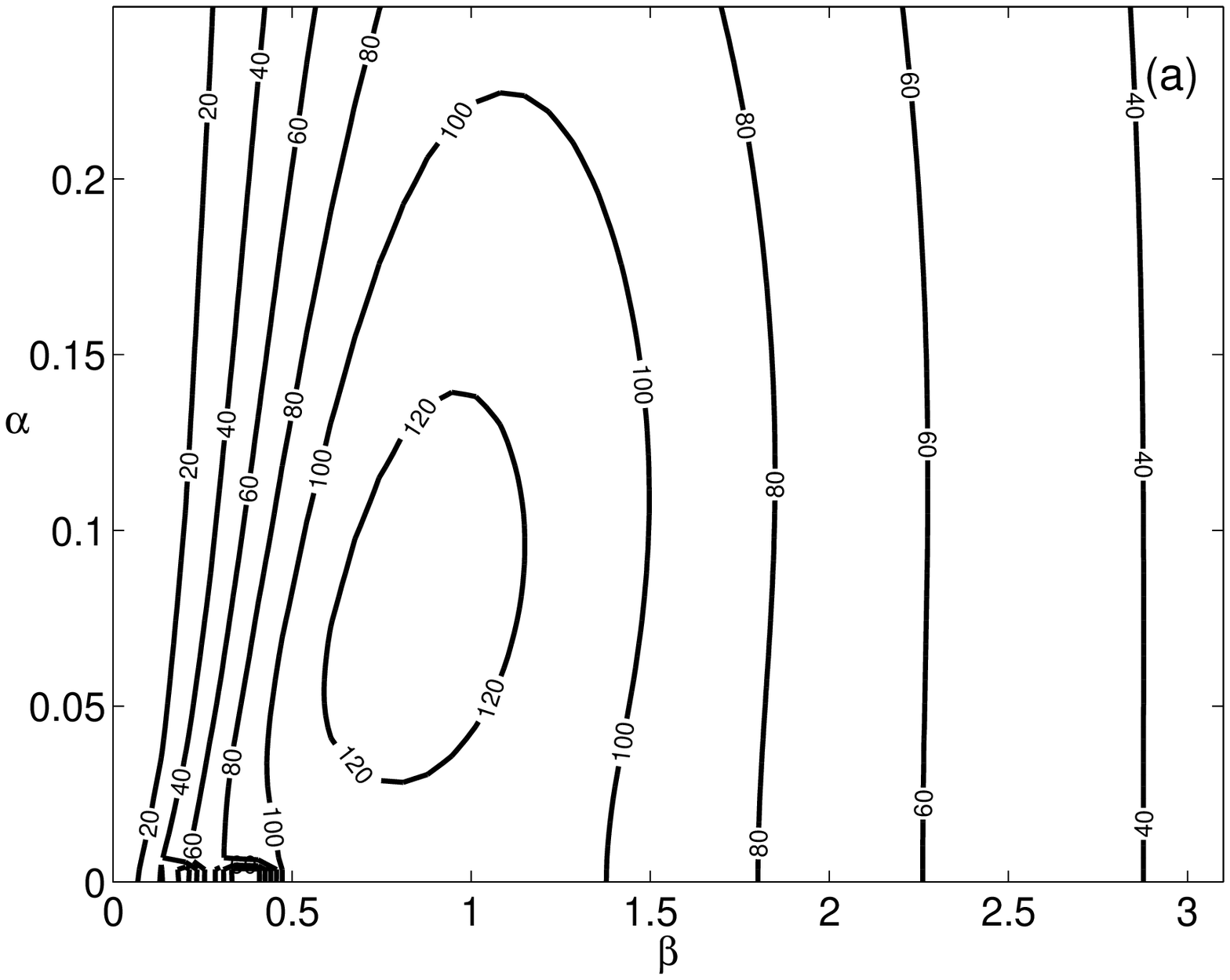}

  \includegraphics[width=3.0in]{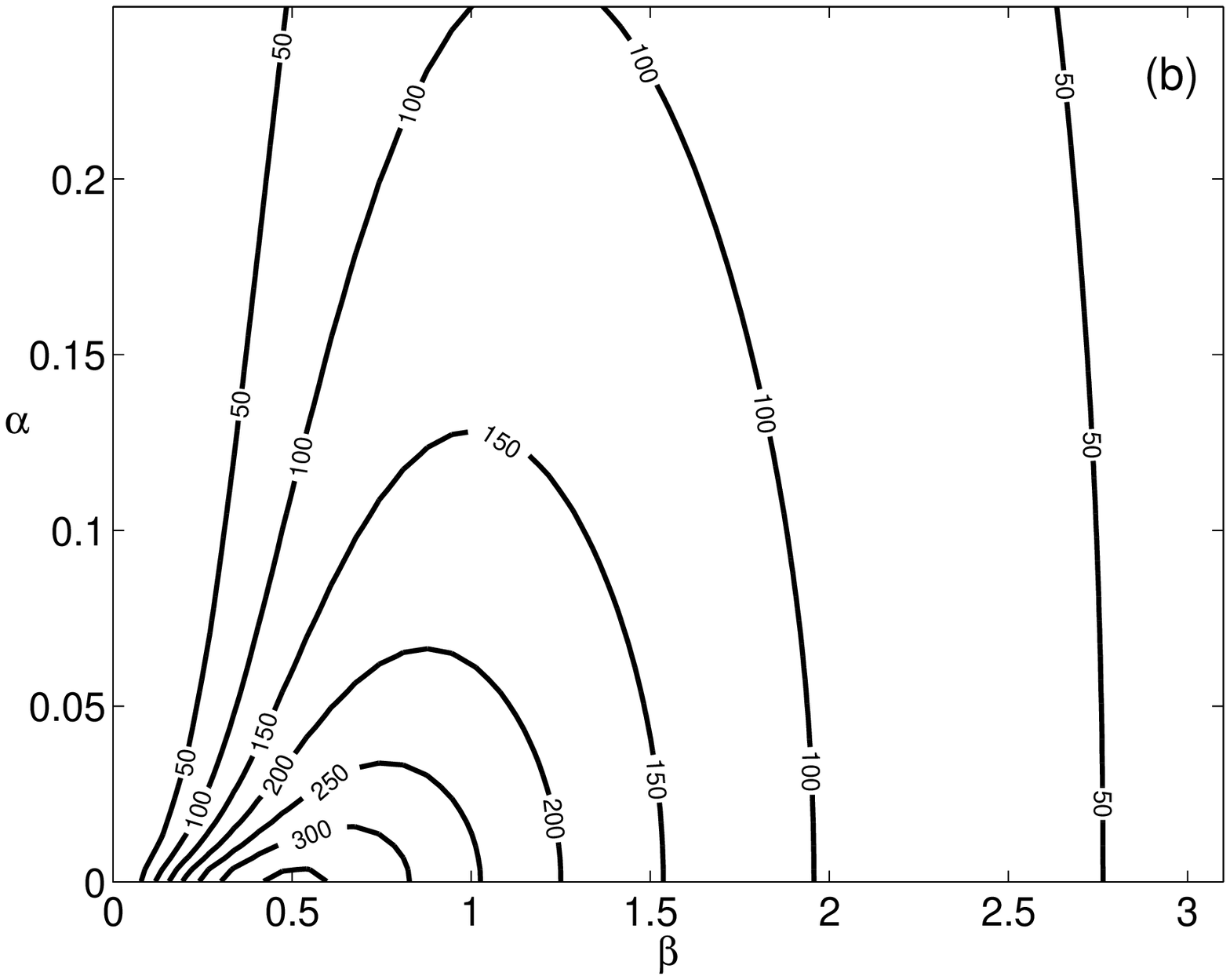}

  \includegraphics[width=3.0in]{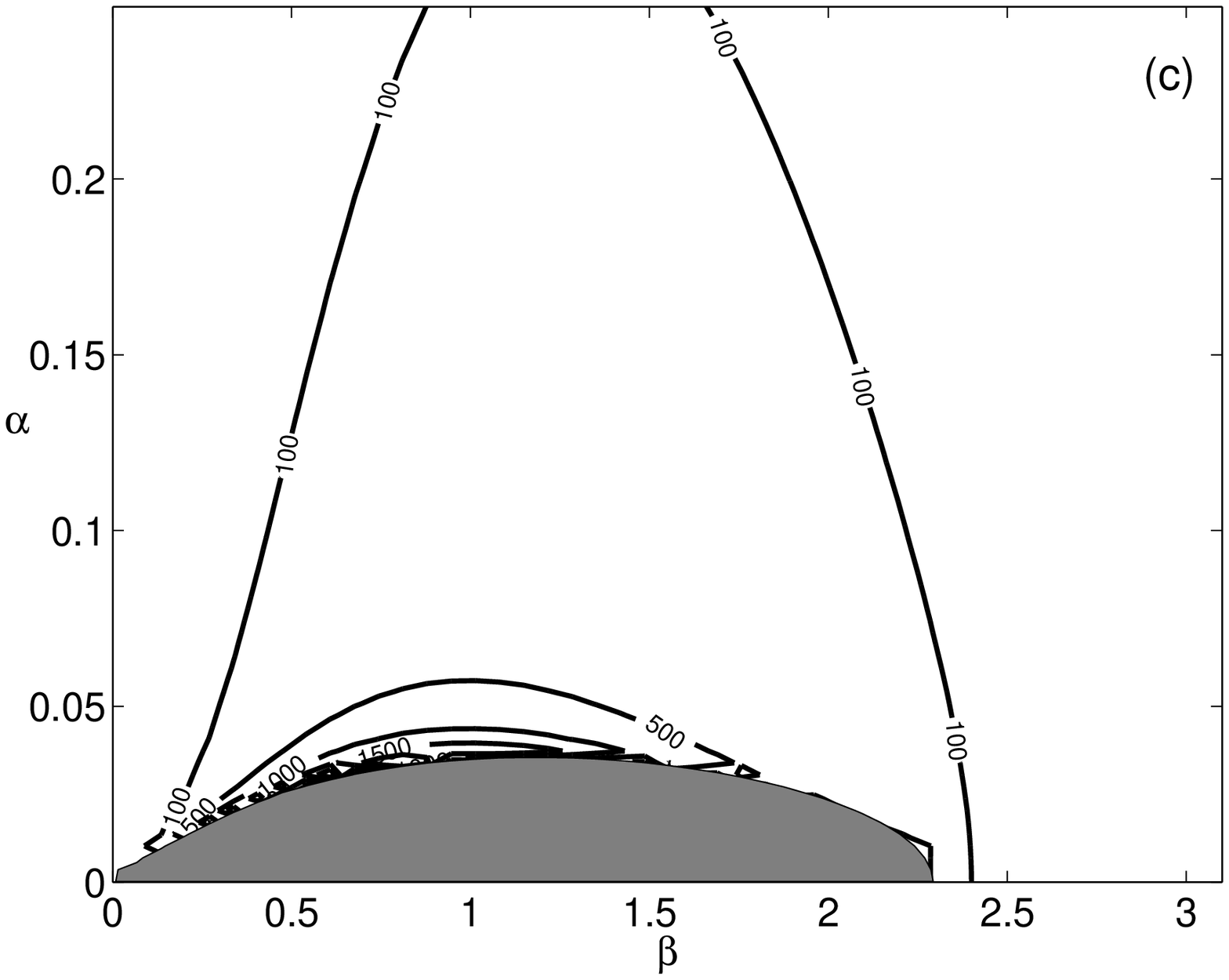}
\end{center}
\caption{Yecko, Phys. Fluids}
\label{rsuction3fig}
\end{figure*}

\clearpage
\newpage
\begin{figure*}
\begin{center}
  \includegraphics[width=3.25in]{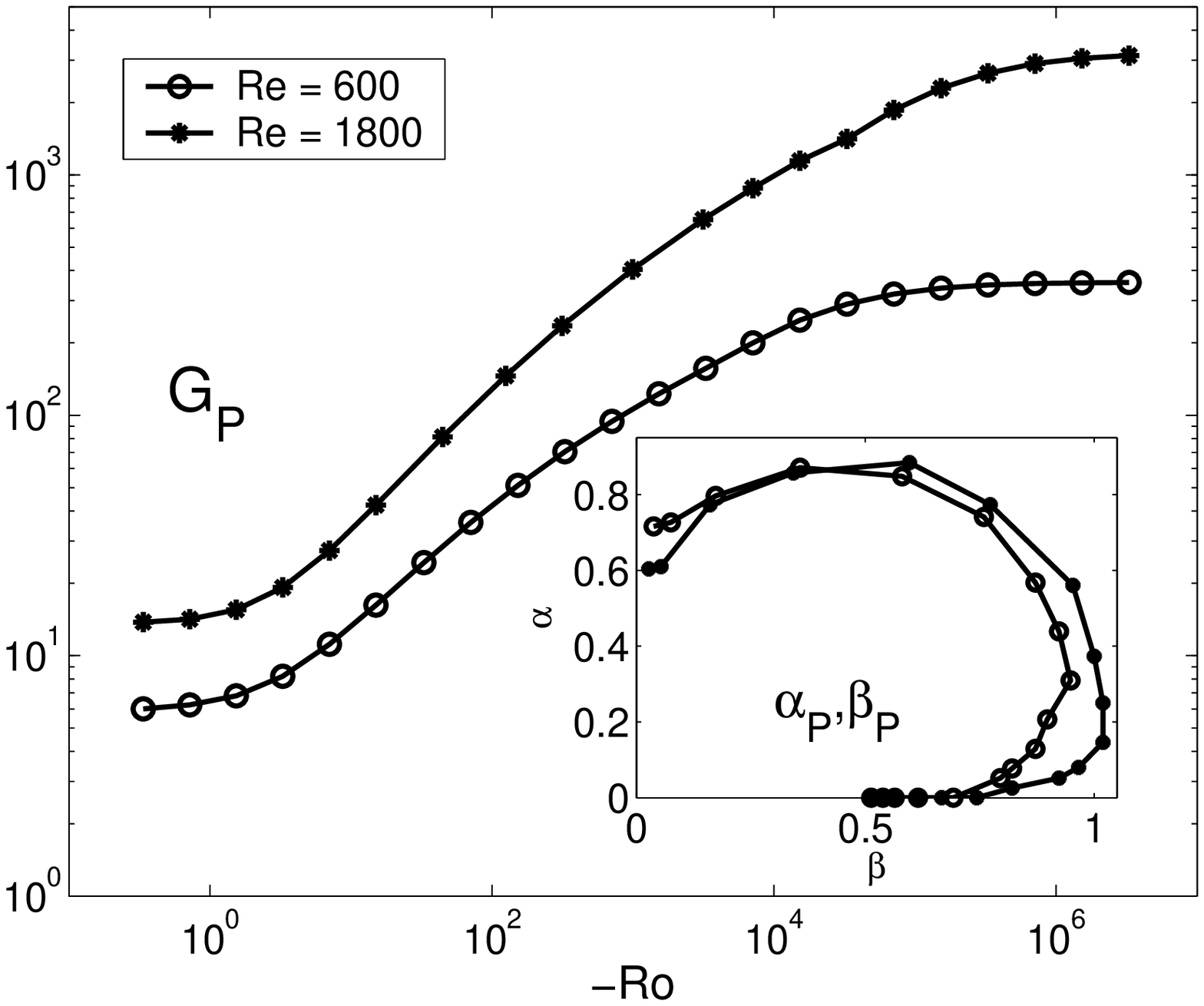}
\end{center}
\caption{Yecko, Phys. Fluids}
\end{figure*}

\clearpage
\newpage
\begin{figure*}
\begin{center}
  \includegraphics[width=3.000in]{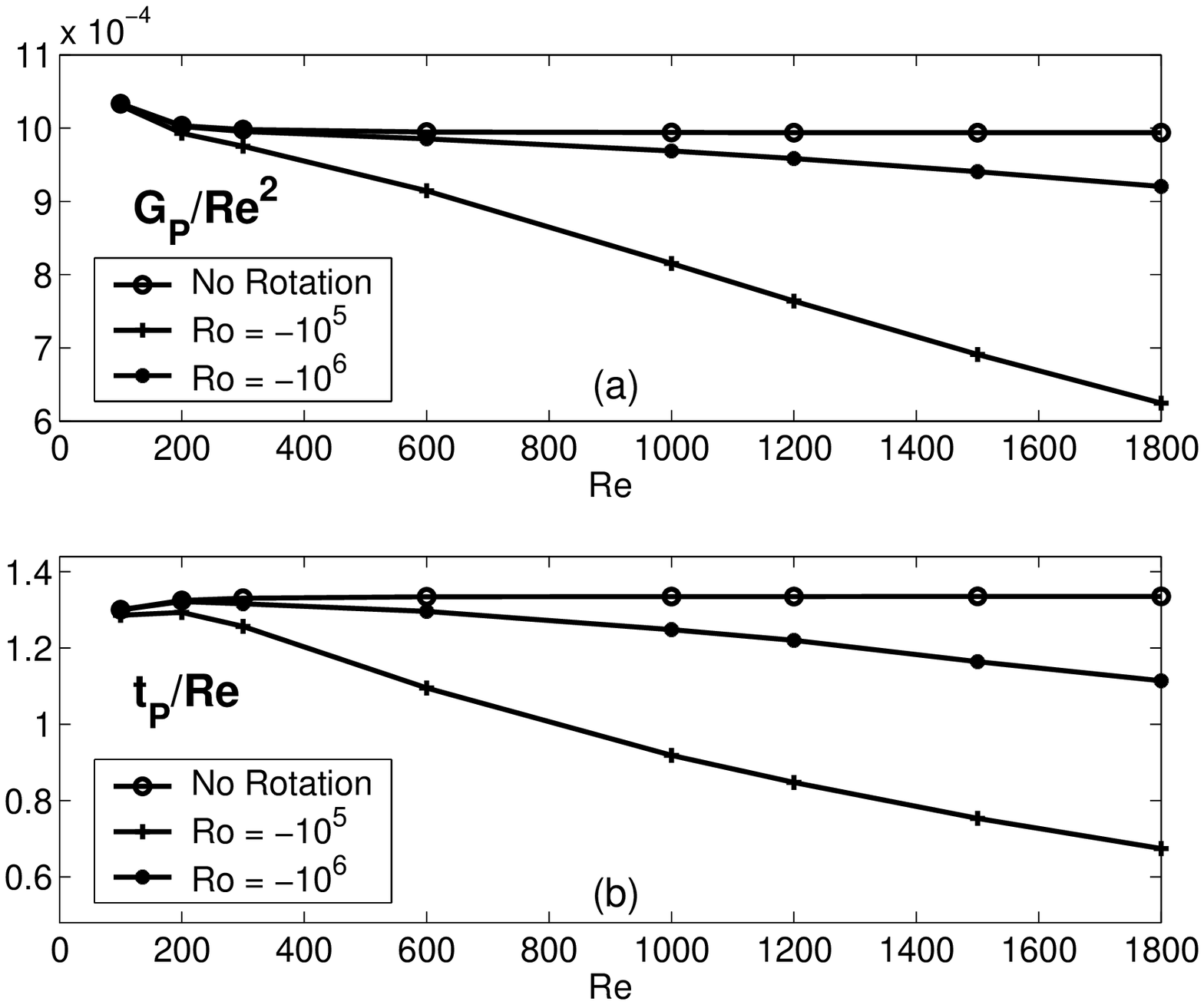}
  \includegraphics[width=3.000in]{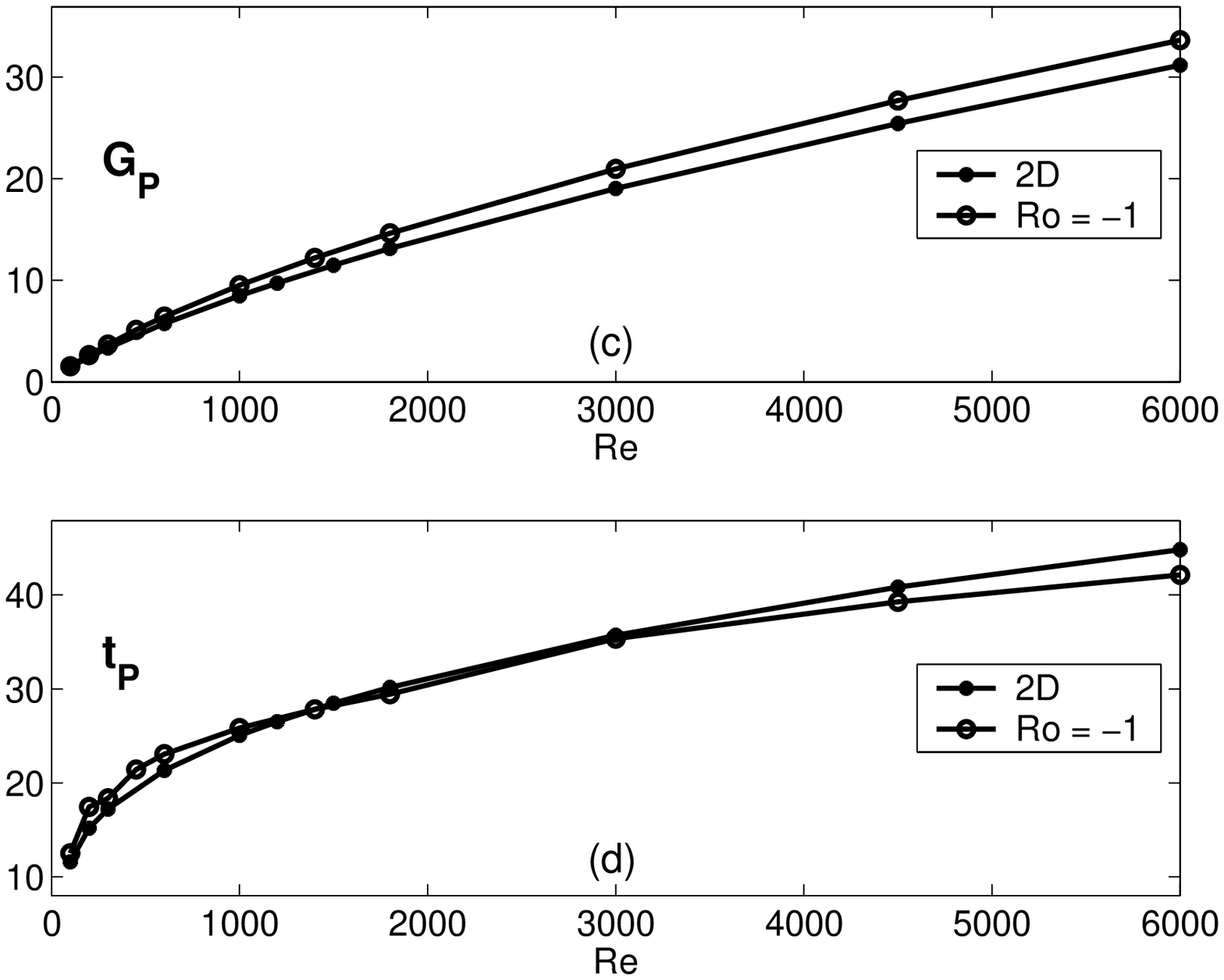}
\end{center}
\caption{Yecko, Phys. Fluids}
\label{fig10}
\end{figure*}

\clearpage
\newpage
\begin{figure*}
\begin{center}
  \includegraphics[width=2.750in]{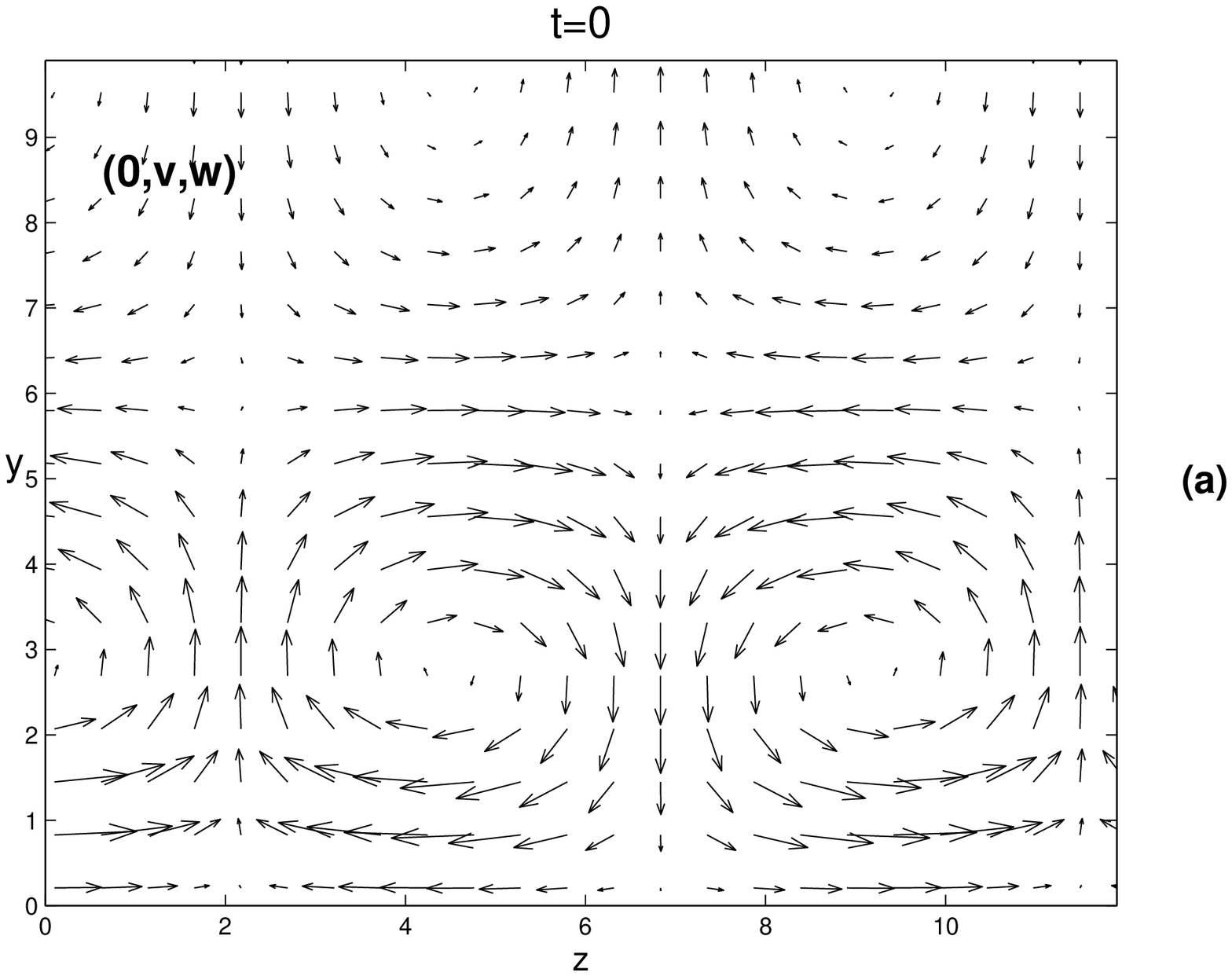}
  \includegraphics[width=2.500in]{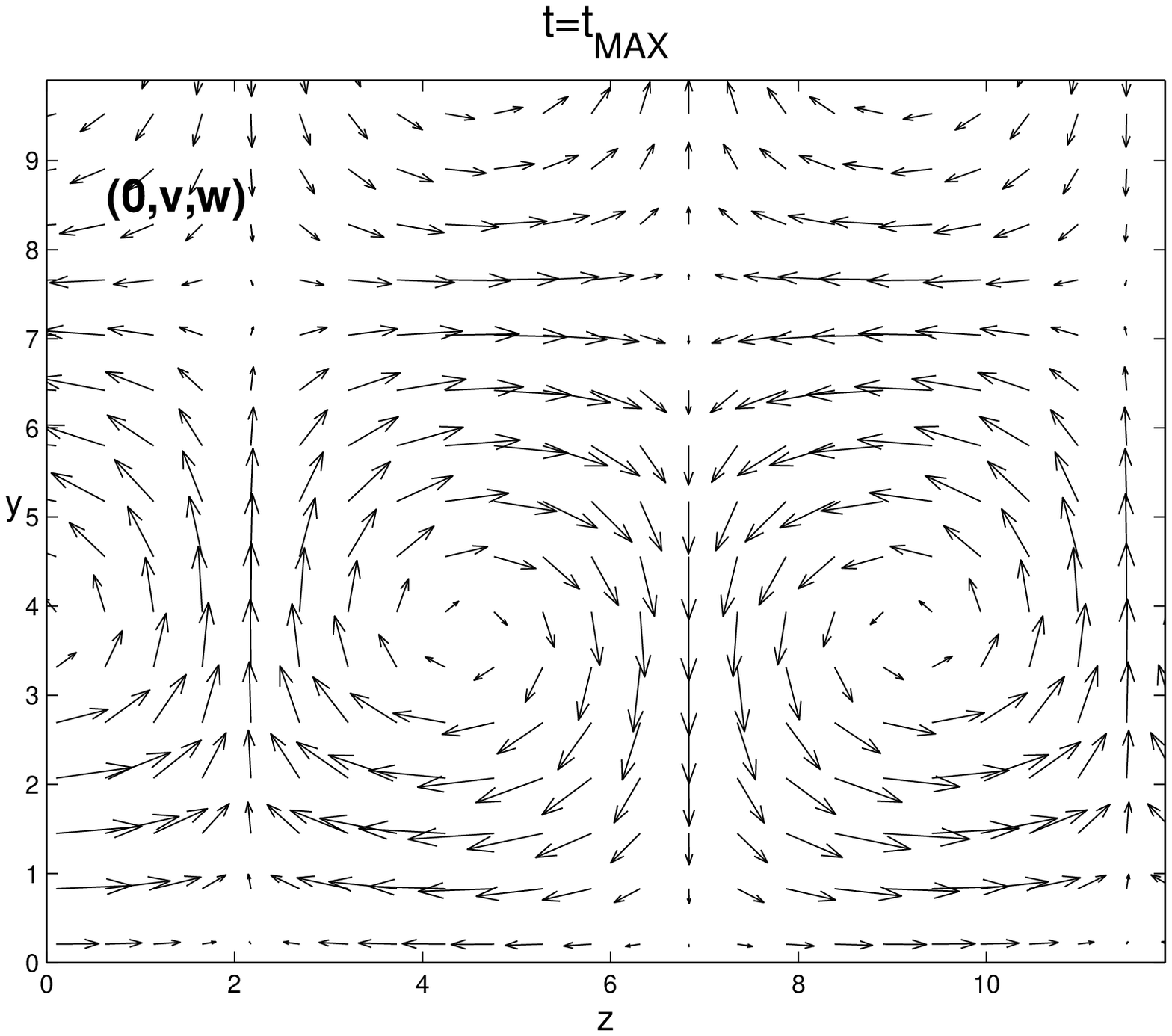}

  \includegraphics[width=2.750in]{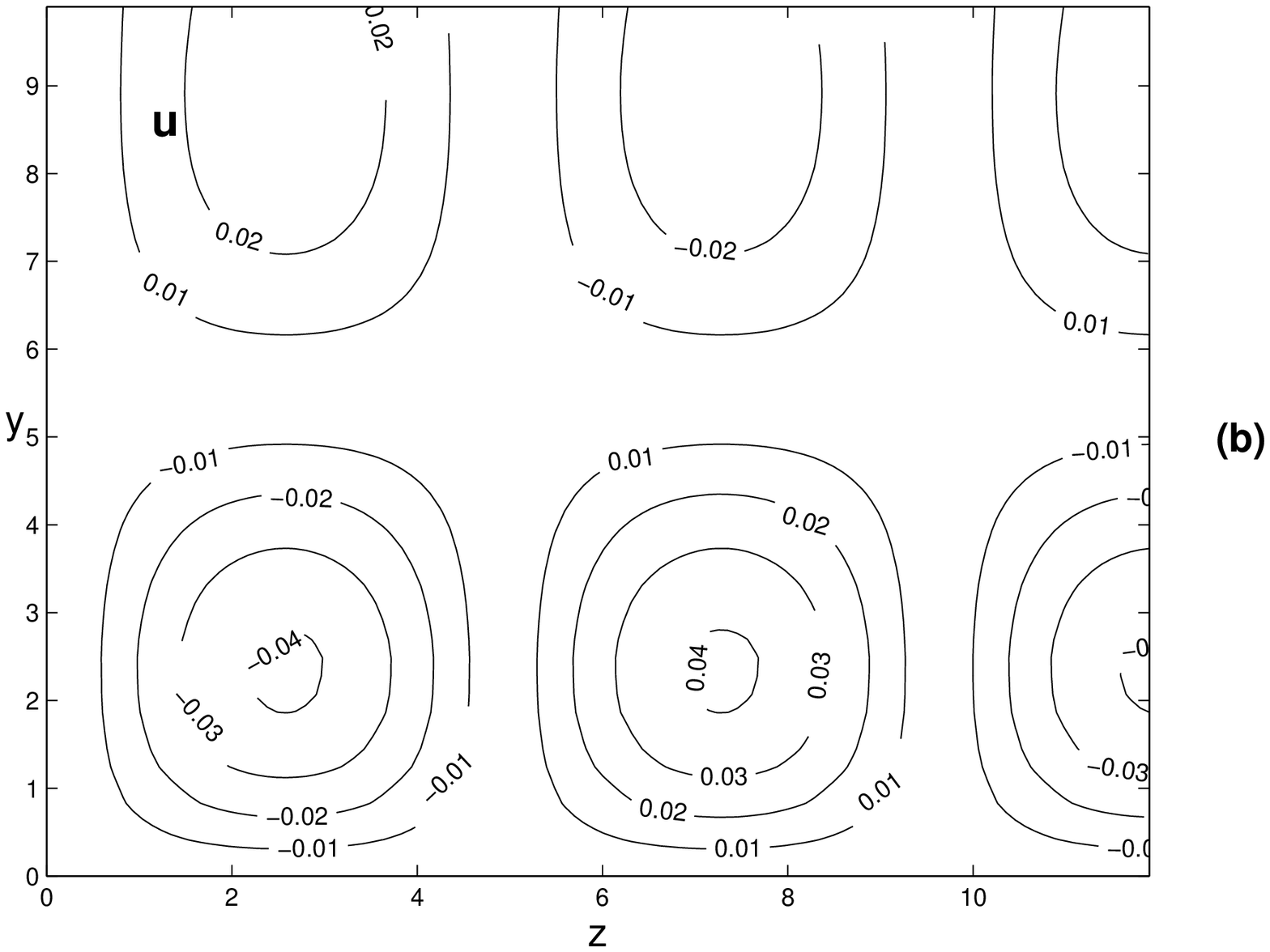}
  \includegraphics[width=2.500in]{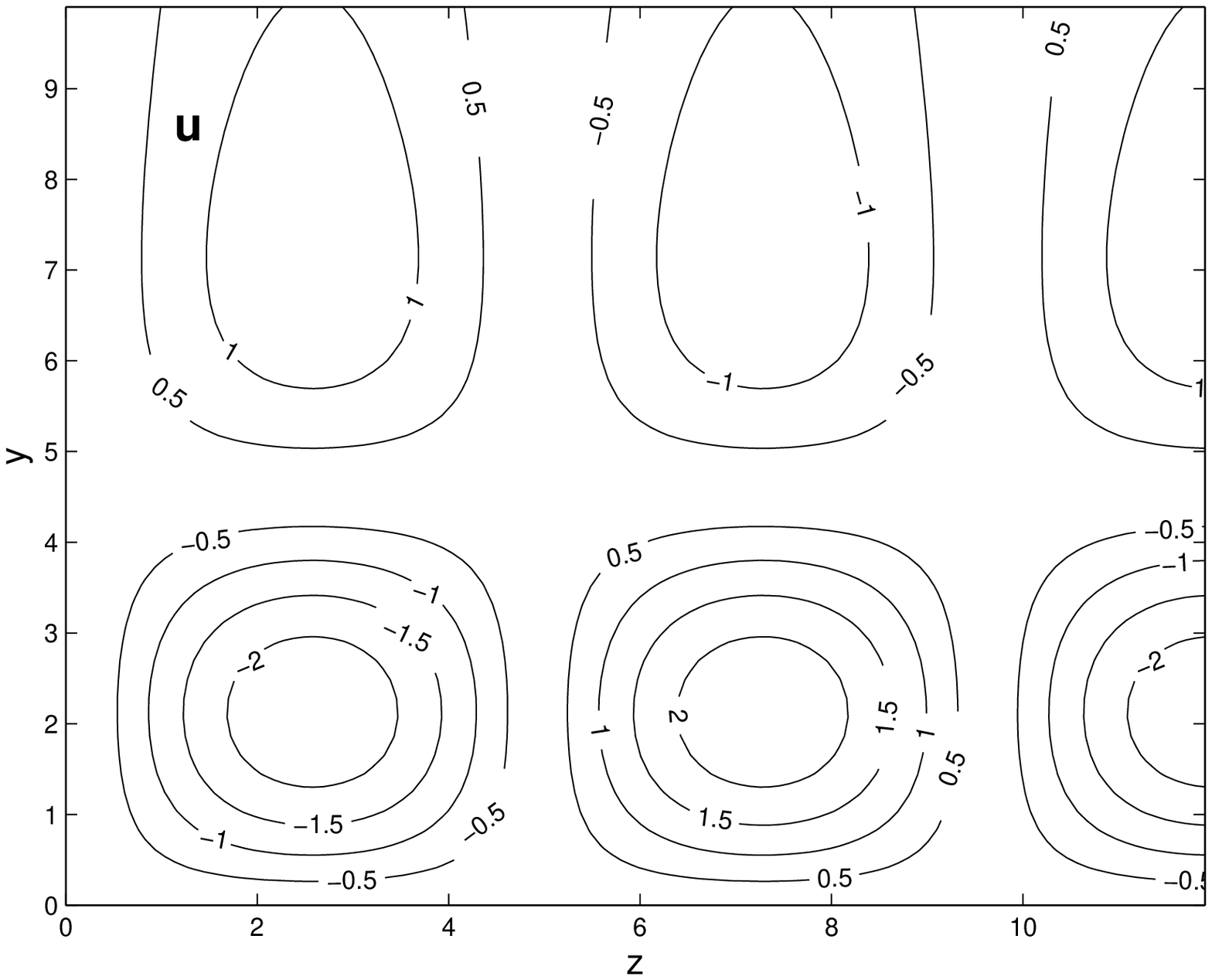}

  \includegraphics[width=2.750in]{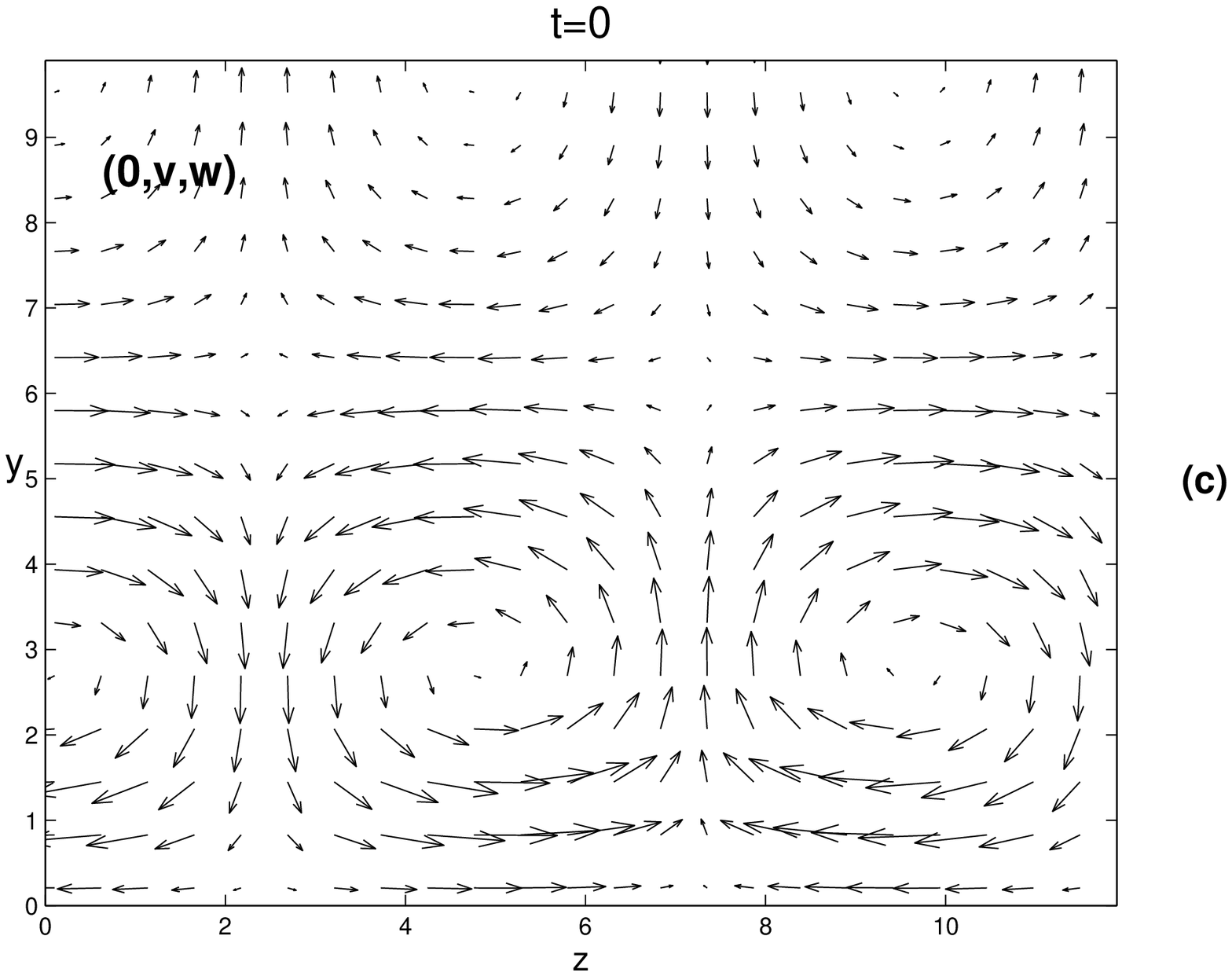}
  \includegraphics[width=2.500in]{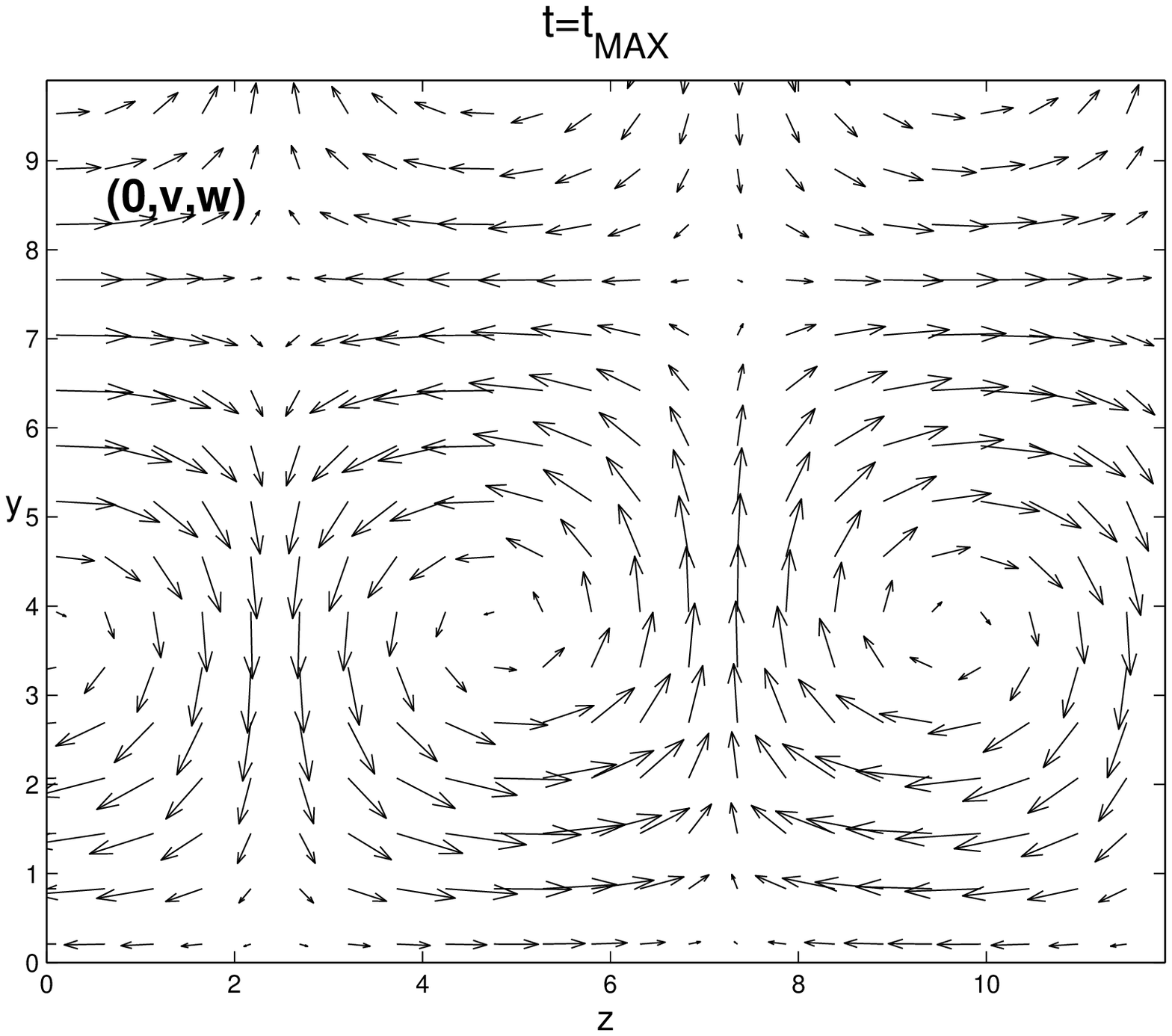}

  \includegraphics[width=2.750in]{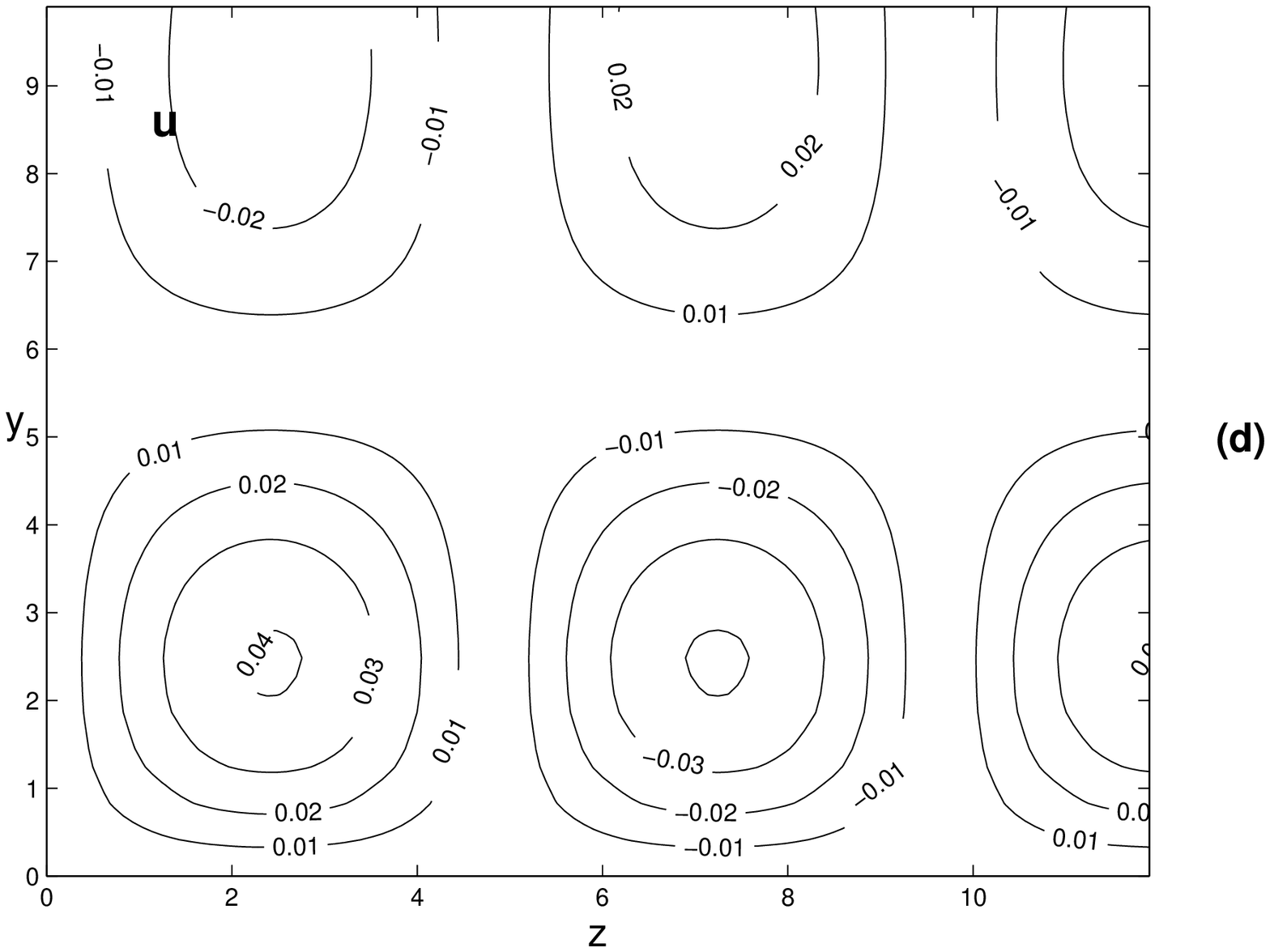}
  \includegraphics[width=2.500in]{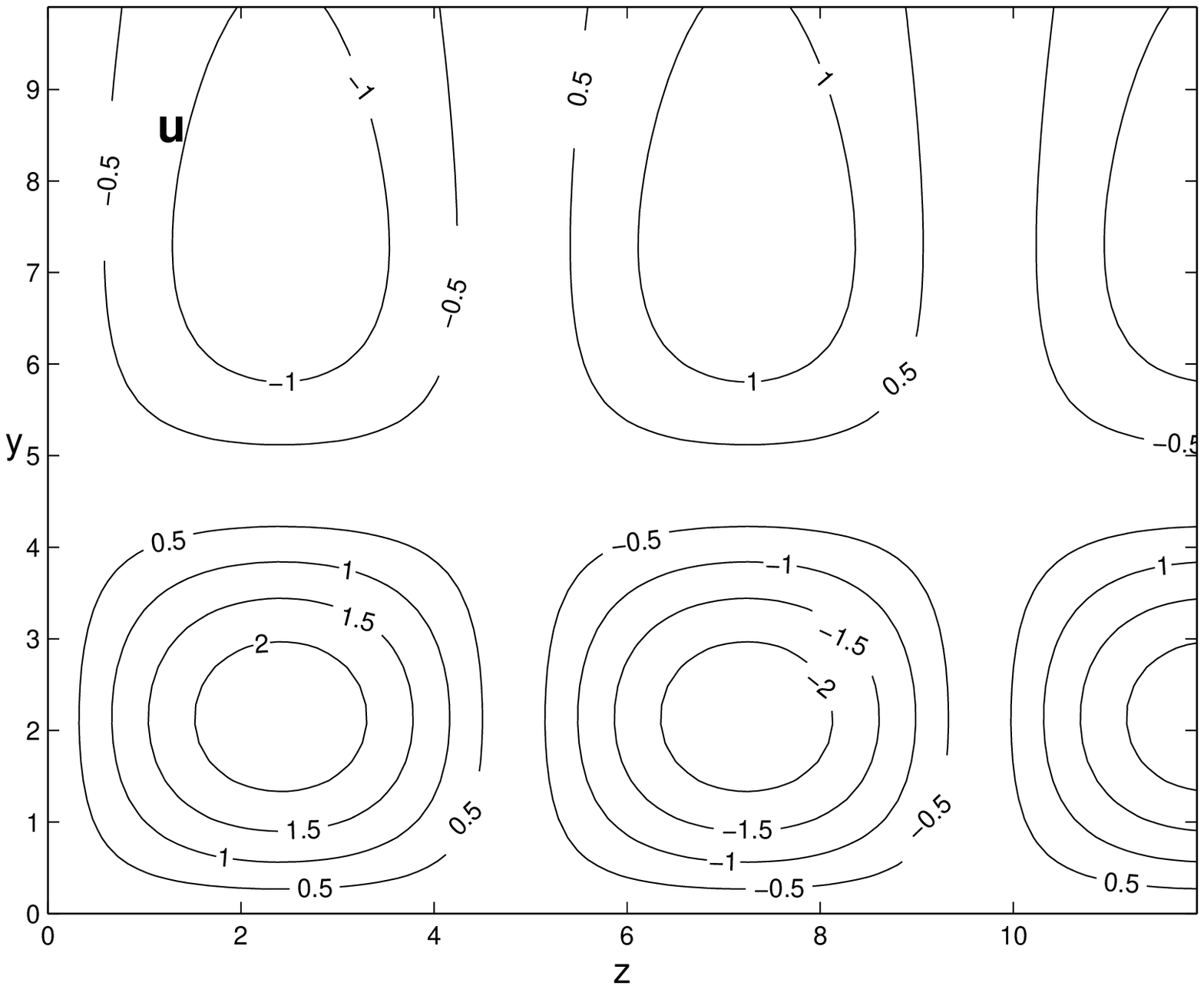}
\end{center}
\caption{Yecko, Phys. Fluids}
\label{optidist}
\end{figure*}

\clearpage
\newpage
\begin{figure*}
\begin{center}
  \includegraphics[width=2.750in]{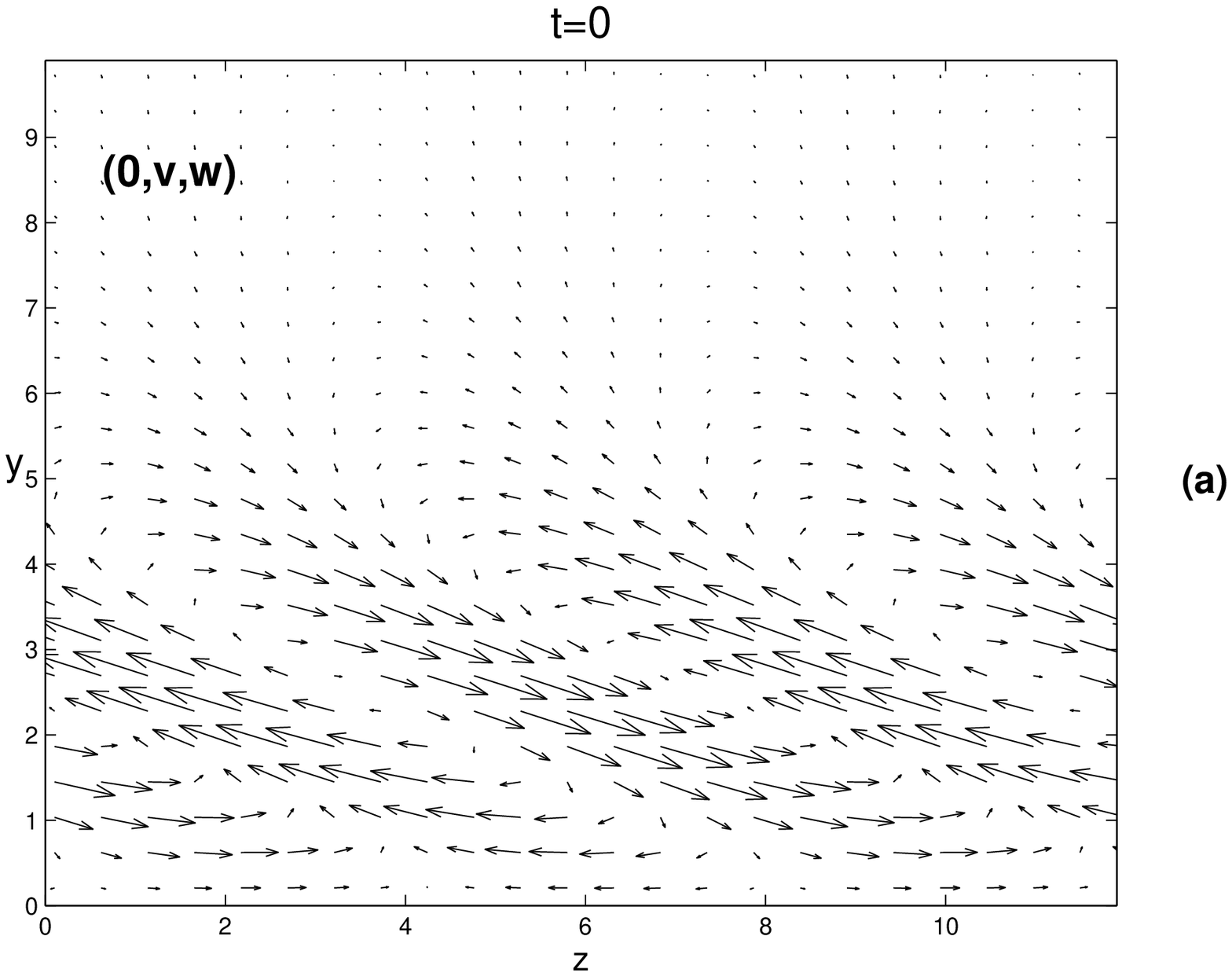}
  \includegraphics[width=2.500in]{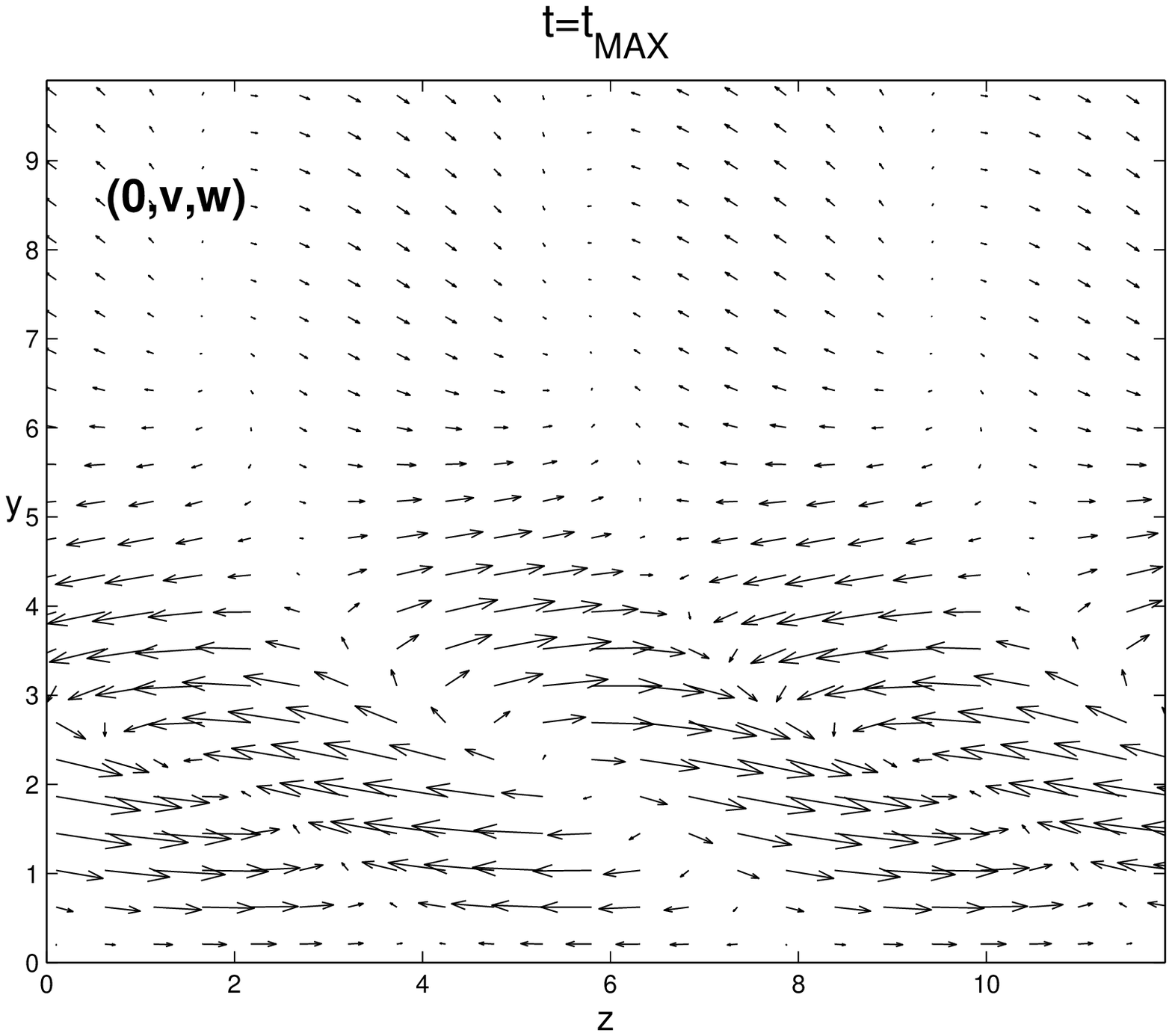}
 
  \includegraphics[width=2.750in]{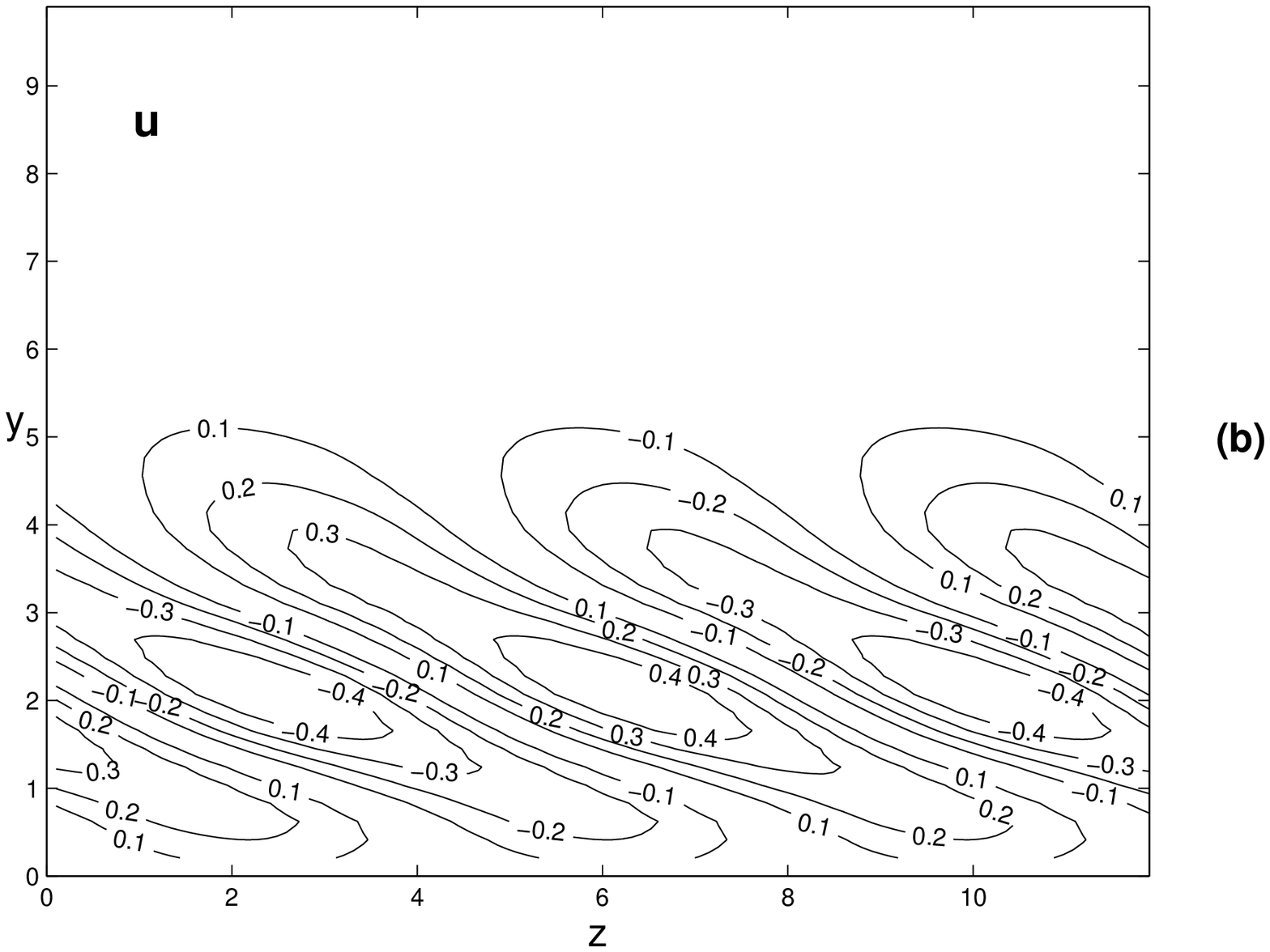}
  \includegraphics[width=2.500in]{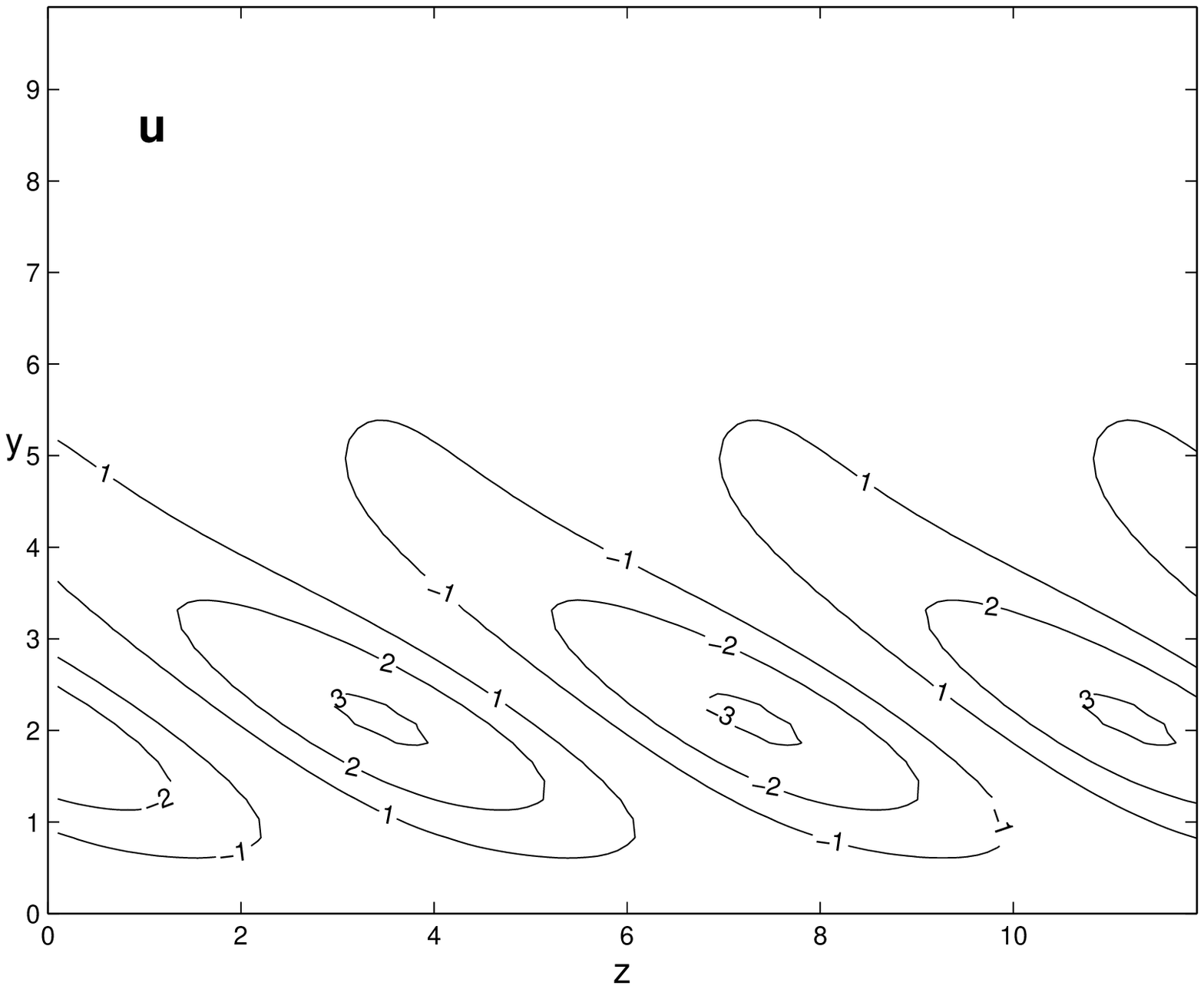}
\end{center}
\caption{Yecko, Phys. Fluids}
\label{optidistrot}
\end{figure*}

\clearpage
\newpage
\begin{figure*}
\begin{center}
  \includegraphics[width=3.000in]{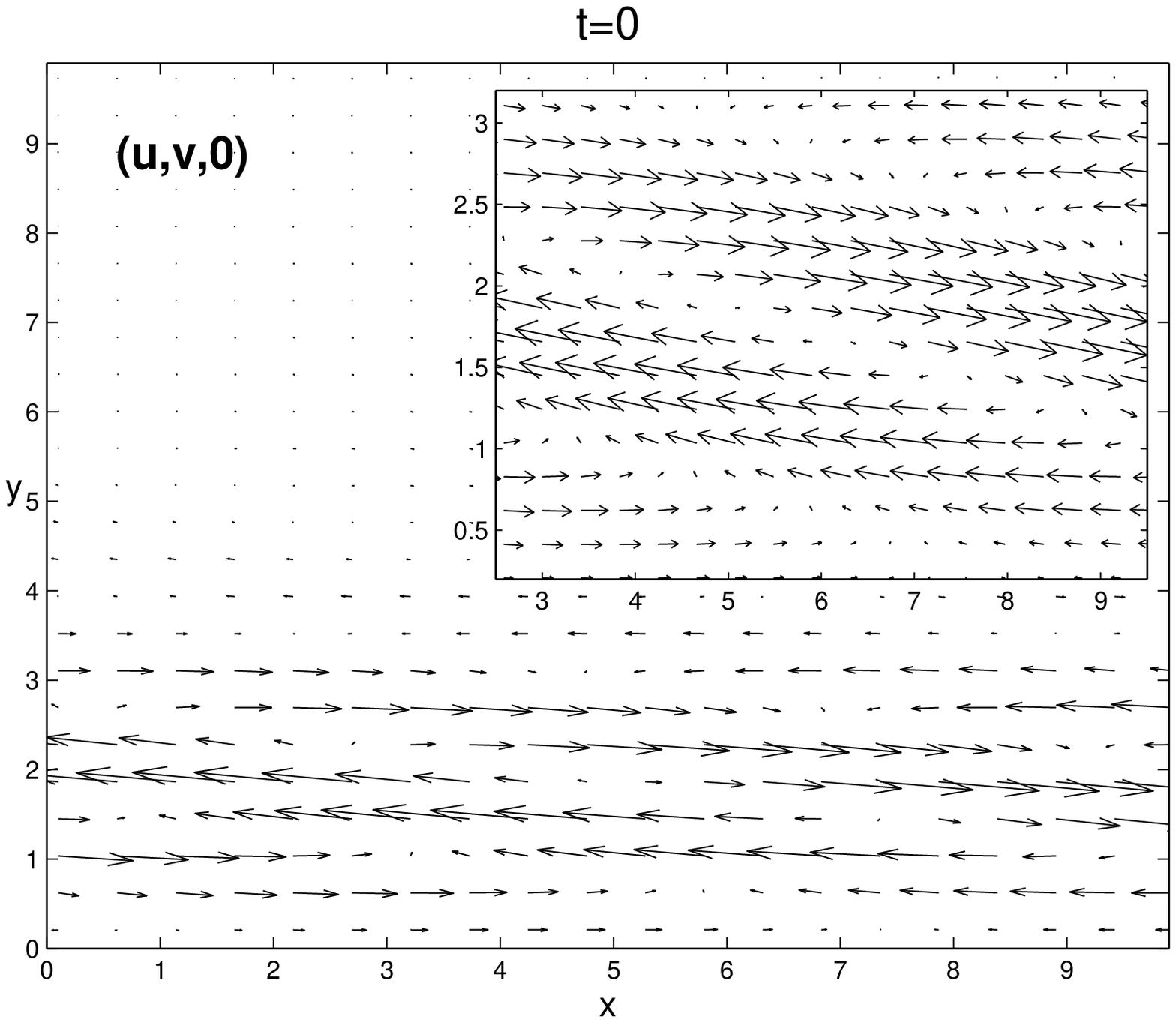}
  \includegraphics[width=3.000in]{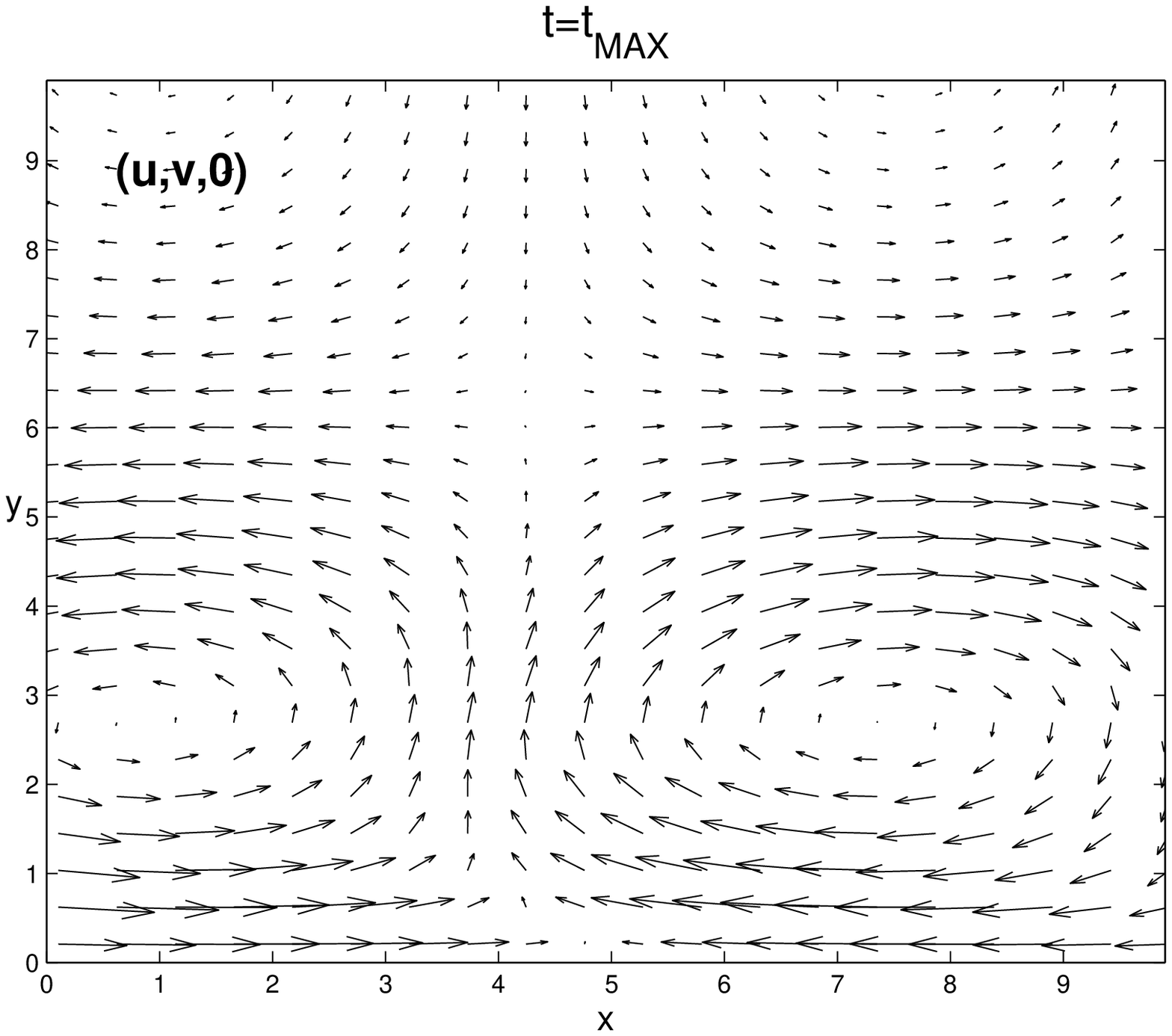}
\end{center}
\caption{Yecko, Phys. Fluids}
\label{optidist2d}
\end{figure*}

\clearpage
\newpage
\begin{figure*}
\begin{center}
  \includegraphics[width=3.250in]{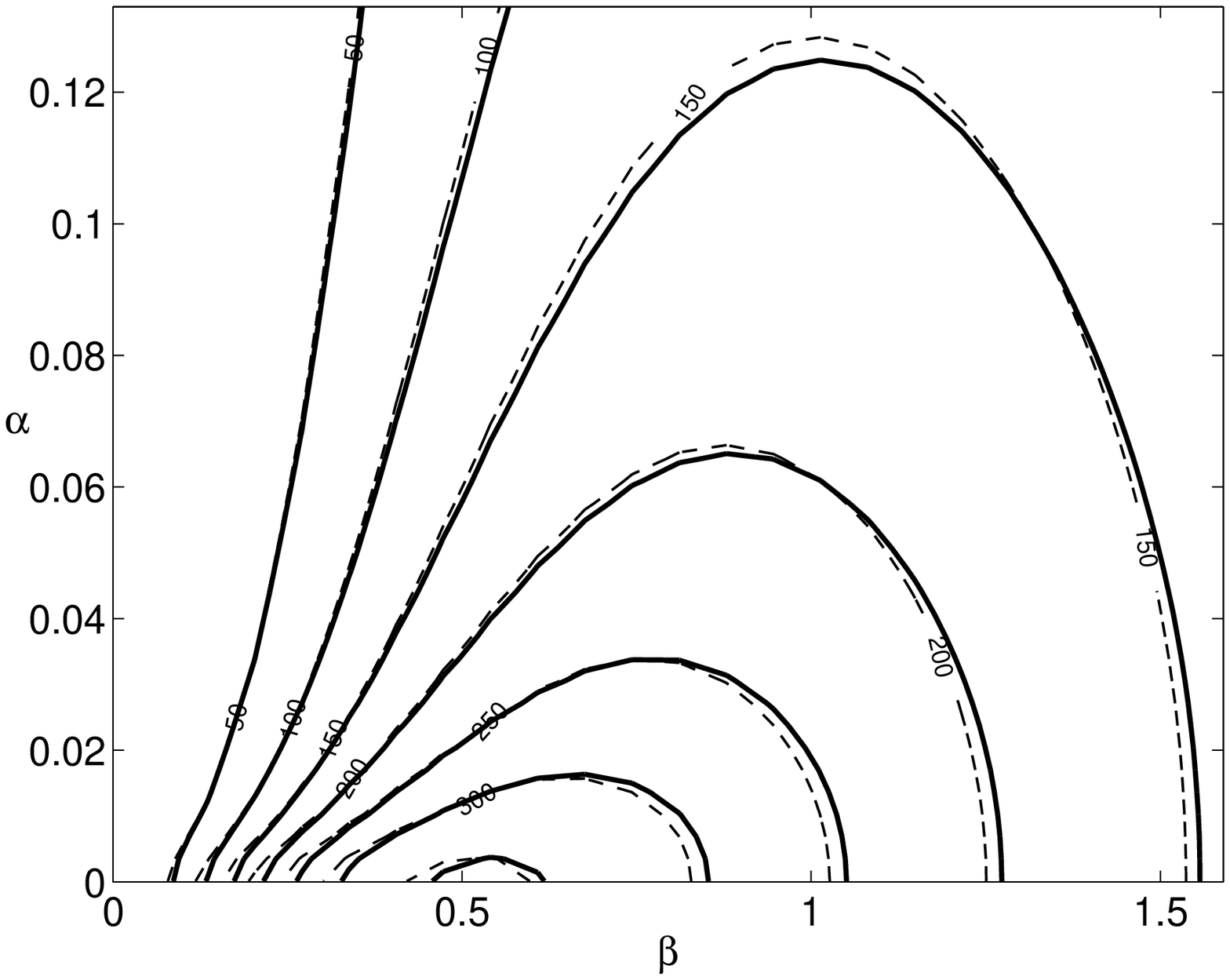}
\end{center}
\caption{Yecko, Phys. Fluids}
\end{figure*}

\clearpage
\newpage
\begin{figure*}
\begin{center}
  \includegraphics[width=3.000in]{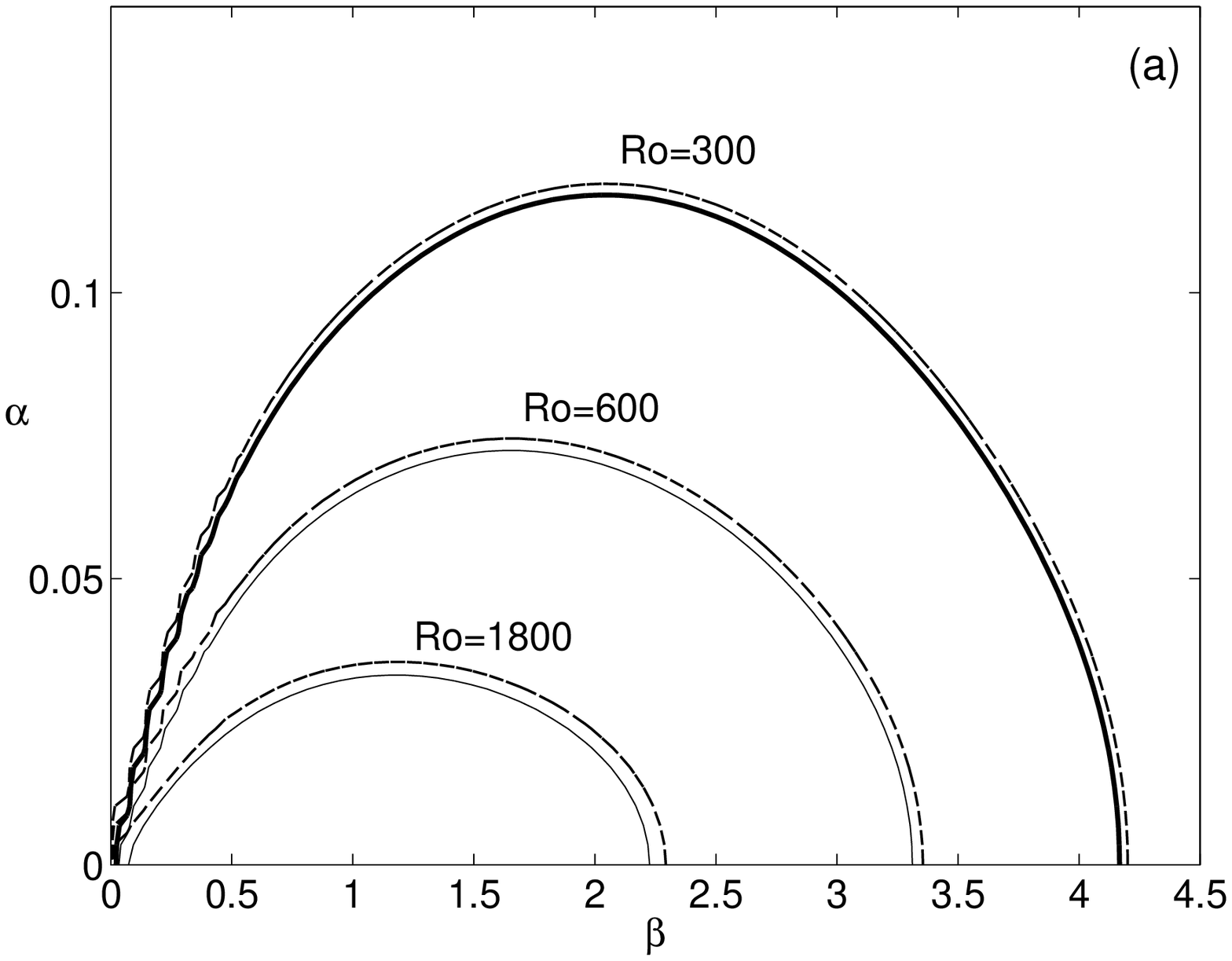}
  \includegraphics[width=3.000in]{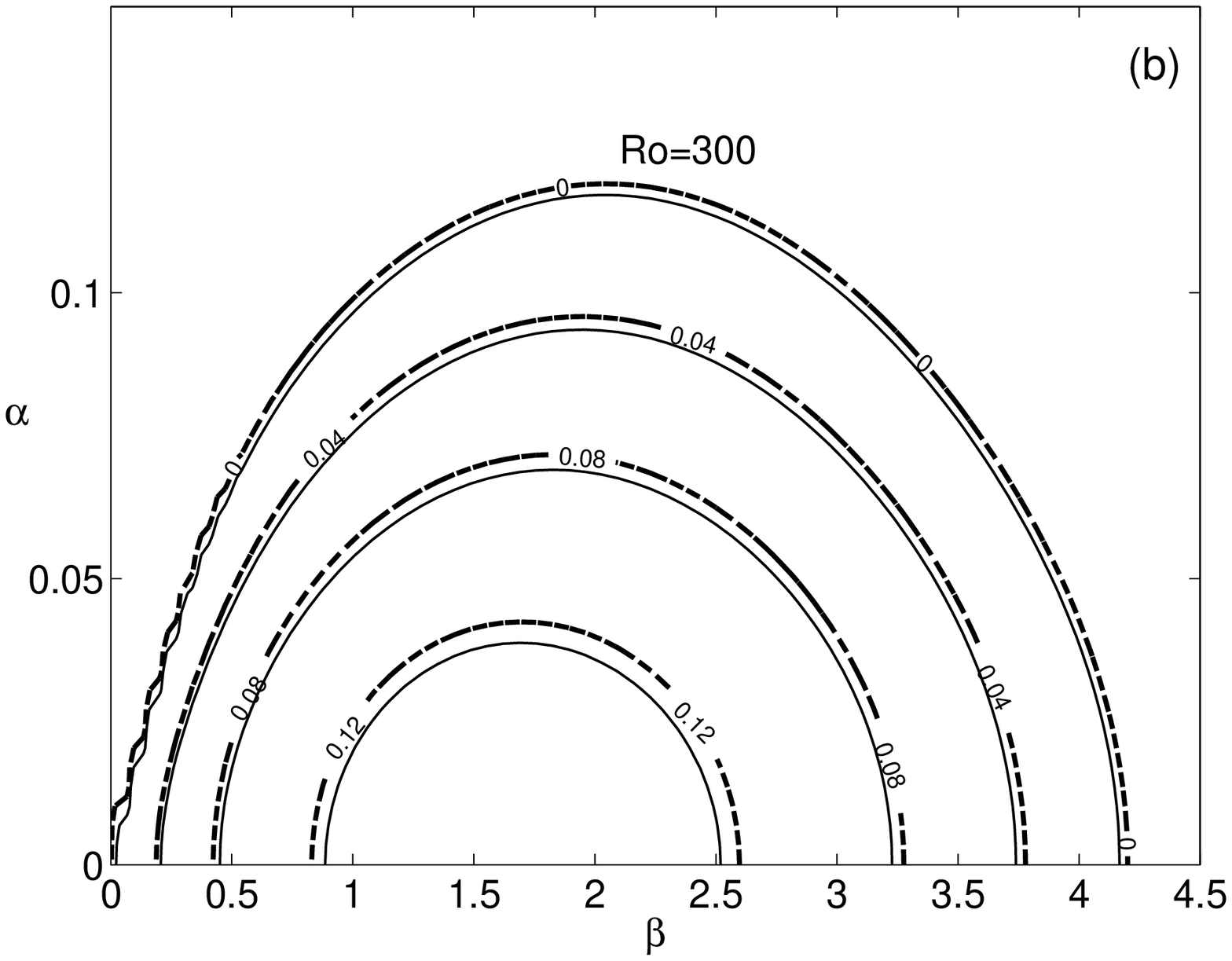}
\end{center}
\caption{Yecko, Phys. Fluids}
\end{figure*}

\clearpage
\newpage
\begin{figure*}
\begin{center}
  \includegraphics[width=3.250in]{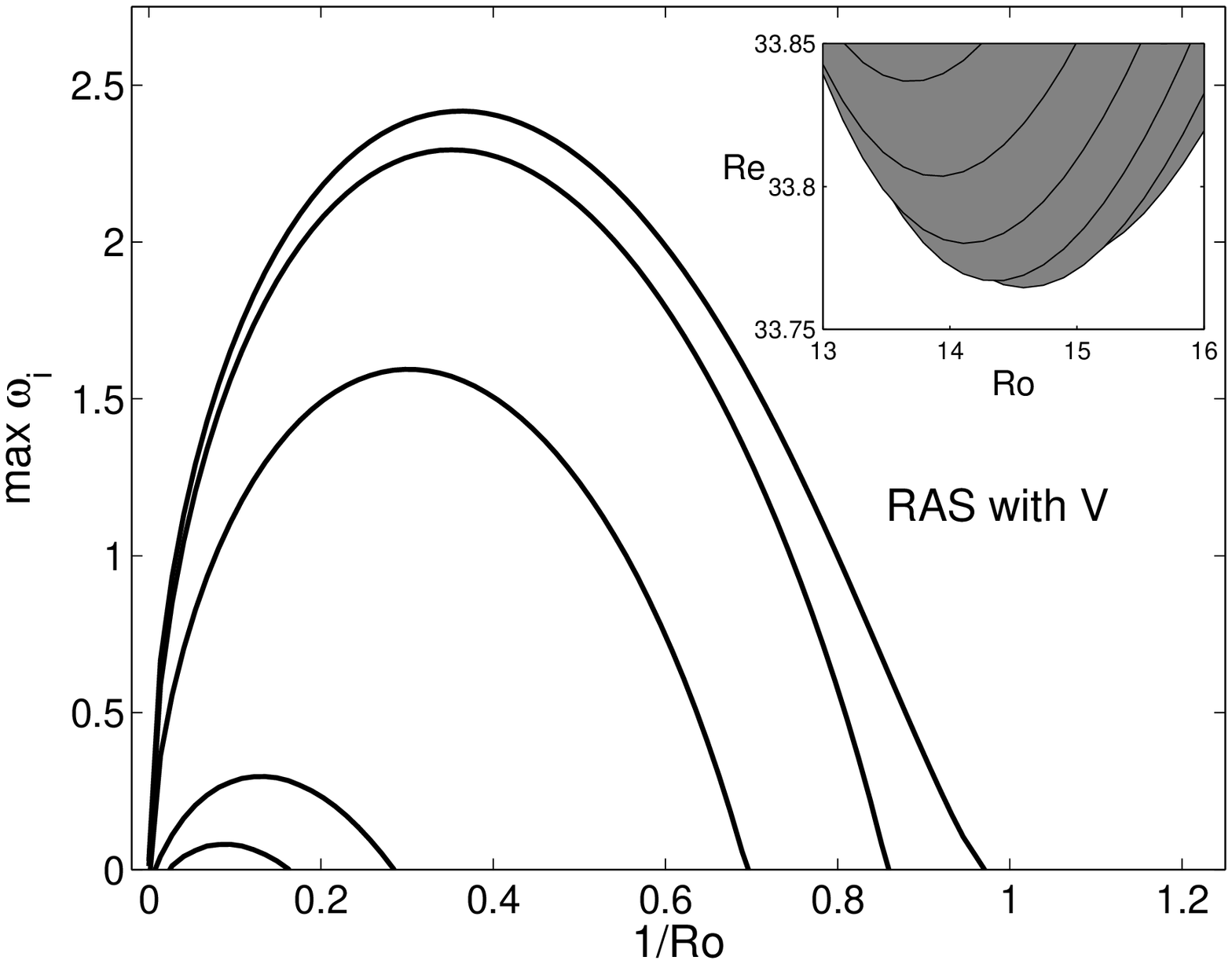}
\end{center}
\caption{Yecko, Phys. Fluids}
\end{figure*}

\clearpage
\newpage
\begin{figure*}
\begin{center}
  \includegraphics[width=3.250in]{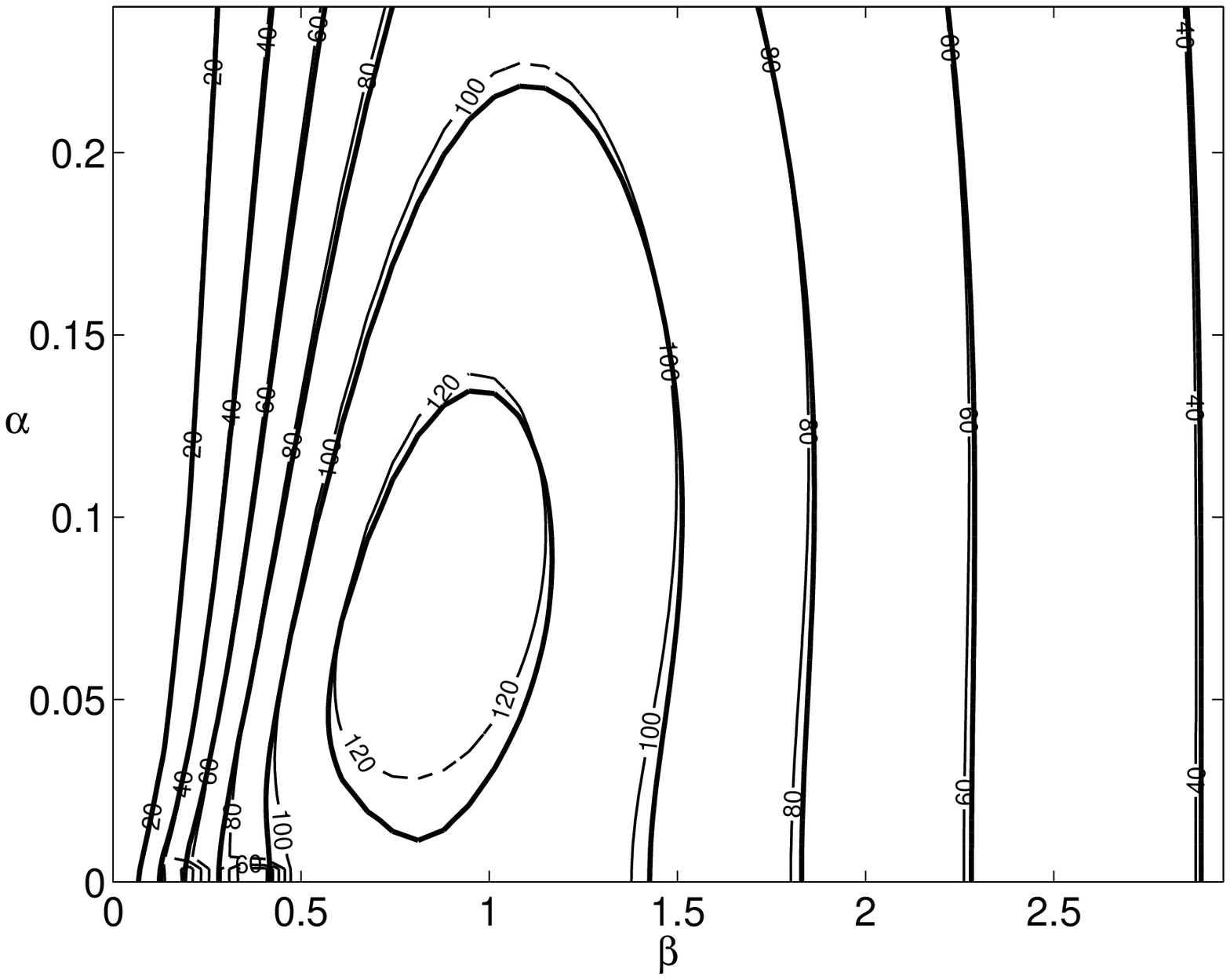}
\end{center}
\caption{Yecko, Phys. Fluids}
\end{figure*}


\begin{thebibliography}{}
\bibitem{footnote}
A constant
transverse base velocity is included in this discussion to include the RAS flow with $V \ne 0$.
In the other cases, it will be turned off i.e. $V=0$.
\end{thebibliography}
\end{document}